\documentclass[12pt,letterpaper]{article}
 
\usepackage{amsfonts}

\usepackage{graphicx}

\def\be{\begin{equation}}
\def\ee{\end{equation}}
\def\bear{\begin{eqnarray}}
\def\eear{\end{eqnarray}}
\def\nn{\nonumber}

\begin{document}

%%%%%%%%%%%%%%%%%%%%%%%%%%%%%%%%%%%%%%%%%%%%%%%%%%%%%%%%%%%%%%%%%%%%
%  TITLE PAGE                                                      %
%%%%%%%%%%%%%%%%%%%%%%%%%%%%%%%%%%%%%%%%%%%%%%%%%%%%%%%%%%%%%%%%%%%%
%  Find all ***                                                    %
%%%%%%%%%%%%%%%%%%%%%%%%%%%%%%%%%%%%%%%%%%%%%%%%%%%%%%%%%%%%%%%%%%%%

\begin{titlepage}
%\begin{flushright}
%yymm.nnnn
%\end{flushright}
\vskip 1in
\begin{center}
{\Large
{Holographic entanglement entropy in nonlocal theories}}
\vskip 0.5in
{Joanna L. Karczmarek and Charles Rabideau}
\vskip 0.3in
{\it 
Department of Physics and Astronomy\\
University of British Columbia,
Vancouver, Canada}
\end{center}

\vskip 0.5in
\begin{abstract}
We compute holographic entanglement entropy in two strongly coupled
nonlocal field theories: the dipole and the noncommutative
deformations of SYM theory.  We find that entanglement entropy  
in the dipole theory follows a volume law
for regions smaller than the length scale of nonlocality and
has a smooth cross-over to an area law for larger regions.  In
contrast, in the noncommutative theory the entanglement entropy
follows a volume law for up to a critical
length scale at which a phase transition to
an area law occurs.  The critical length scale
increases as the UV cutoff is raised, which is
indicative of UV/IR mixing 
and implies that entanglement entropy in the noncommutative theory
follows a volume law
for arbitrary large regions when the size of the region is
fixed as the UV cutoff is removed to infinity.
Comparison of behaviour between these two theories allows us to explain the origin of the volume law.
Since our holographic duals  are not asymptotically AdS, 
minimal area surfaces used to compute holographic entanglement entropy
have novel behaviours near the boundary of the dual spacetime.
We discuss implications of our results on the 
scrambling (thermalization) behaviour of these nonlocal field theories. 
\end{abstract}
\end{titlepage}

\tableofcontents

%%%%%%%%%%%%%%%%%%%%%%%%%%%%%%%%%%%%%%%%%%%%%%%%%%%%%%%%%%%%%%%%%%%%
%  DRAFTMORE ONLY; ADD MORE AUTHORS IF NEEDED                      %
%%%%%%%%%%%%%%%%%%%%%%%%%%%%%%%%%%%%%%%%%%%%%%%%%%%%%%%%%%%%%%%%%%%%

%\begin{flushright} {\small {Printout date: \today}} \end{flushright}
%Notes in progress by Joanna Karczmarek.

% replaced by options in the headings

%%%%%%%%%%%%%%%%%%%%%%%%%%%%%%%%%%%%%%%%%%%%%%%%%%%%%%%%%%%%%%%%%%%%
%  BEGIN HERE                                                      %
%%%%%%%%%%%%%%%%%%%%%%%%%%%%%%%%%%%%%%%%%%%%%%%%%%%%%%%%%%%%%%%%%%%%

\section{Introduction}
\label{sec:intro}

Geometric entanglement entropy as a tool to characterize physical
properties of quantum field theories has recently received a large
amount of attention. 
One attractive feature of geometric entanglement
entropy as an observable is that it is defined in the same way
in any quantum field theory: it is simply the von Neumann entropy,
$-\mathrm{Tr}(\rho_A \log \rho_A)$,
associated with the density matrix $\rho_A$ describing 
degrees of freedom living 
inside a region $A$.  $\rho_A$ arises when the portion of total Hilbert space 
associated with degrees of freedom living outside of $A$ is traced over.
Universality of entanglement entropy is reflected in the
Ryu-Takayanagi holographic formula \cite{Ryu:2006bv}
\be
S[A] = \frac{\mathrm{Vol_d}(\tilde A)} {4G_N^{(d+2)}}~.
\label{RT-conformal}
\ee
Here, we place $A$, a d-dimensional spacial region, on a spacelike slice of
the boundary of the (d+2)-dimensional spacetime dual to the quantum
field theory of interest.  $\tilde A$ is a minimal area surface in the
bulk of the holographic dual spacetime homologous to $A$.
$G_N^{(d+2)}$ is the (d+2)-dimensional Newton constant and the
d-dimensional volume of $\bar A$ is denoted with $\mathrm{Vol_d}(\bar A)$. 
\footnote{For an accessible introduction and some recent developments to
holographic entropy, see for example \cite{Nishioka:2009un,Takayanagi:2012kg}.}

The Ryu-Takayanagi formula (\ref{RT-conformal}) is applicable to
holographic duals where the dilaton and the volume
of the internal sphere are both constant.  However,
duals to the nonlocal theories we are interested in have neither,
so the local gravitational constant $G_N^{(d+2)}$ varies.
Thus we  must use
a generalized version of formula (\ref{RT-conformal}), given by
\cite{Ryu:2006ef}
\be
S[A] = \frac{\mathrm{Vol}(\bar A)} {4G_N^{(10)}}  
~,~~~{\mathrm{with}}~
\mathrm{Vol}(\bar A) = \int d^8\sigma e^{-2\phi}
\sqrt{G^{(8)}_{\mathrm{ind}}}~,
\label{RT}
\ee
where  $G^{(10)}_N = 8 \pi^6 (\alpha')^4 g_s^2$
is the (asymptotic) 10-dimensional Newton's constant and $\phi$ is
the local value of the fluctuation in dilaton field 
(so that the local value of the 10-dimensional Newton's constant is 
$G^{(10)}_N e^{2\phi}$).  Integration
is now over a co-dimension two surface $\bar A$ that wraps the
compact internal manifold of the holographic dual.  

Because $\bar A$ wraps the internal manifold, 
its boundary is the direct product of the boundary of $A$,
$\partial A$, and the internal manifold.  
To obtain entanglement entropy, $\bar A$ is chosen to
to have minimal area (we will only work in static spacetimes).
$G^{(8)}_{\mathrm{ind}}$ is the induced string frame
metric on $\bar A$. By considering the
standard relationship between
local Newton's constants in different dimensions:
$G^{(d+2),\mathrm{local}}_N = {G^{(10)}_N e^{2\phi}}/{V_{8-d}}$,
together with $\mathrm{Vol}(\bar A) = V_{8-d}~\mathrm{Vol_d}{\tilde  A}$,
equation (\ref{RT-conformal}) can 
be easily recovered from (\ref{RT}) for a scenario where 
the dilaton is a constant and the internal manifold has a 
constant volume  $V_{8-d}$ (in string metric).
The more general formula (\ref{RT}) has been used 
to study, for example, tachyon condensation
\cite{Nishioka:2006gr} and confinement-deconfinement transition
\cite{Klebanov:2007ws}.  We will refer to the 8-dimensional 
$\mathrm{Vol}(\bar A)$ as the area of the minimal surface from now on.

Generically, geometric entanglement entropy has a UV divergence, so it
needs to be regulated with a UV cutoff.  Holographically, this is accomplished
the usual way by placing the region $A$ on a surface which is removed
from the boundary of the holographic dual spacetime.  Once the
theory has been regulated with a cutoff, geometric entanglement entropy
in the vacuum state can be thought to count effective  degrees 
of freedom inside $A$ that have quantum correlations with
degrees of freedom  outside of $A$.  In other words, it measures the 
the range of quantum correlations generated in the ground state
by the interactions in the Hamiltonian.
For a local theory, degrees of freedom with correlations across the boundary
of $A$ must live near this boundary, which leads to the area law:
entanglement entropy in local theories is generically proportional
to the area of the boundary of $A$, $|\partial A|$.
While the area law has not been proven for a general interacting field theory,
it is expected to generically hold in local theories for the reason
outlined above (see \cite{Eisert:2008ur} for a review, focusing on lattice systems).

In a nonlocal theory, behaviour of entanglement entropy could be expected to
deviate from the area law and this is precisely what we find using holographic
methods at strong coupling. In a simple nonlocal theory with a fixed scale 
of nonlocality $a_L$, a dipole deformation of ${\cal N} = 4$ SYM, we find that 
entanglement entropy is extensive (proportional to the volume of $A$), 
for regions $A$ of size up to $a_L$. At length scales higher than $a_L$, 
it follows an area law, with an effective number of entangled degrees of freedom 
which is proportional to $a_L$.
This is consistent with all degrees of freedom within a region $A$ of size $a_L$ or smaller,
and  not only those living close to the boundary of $A$,
having quantum correlations with the outside of $A$ due to the nonlocal nature of the Hamiltonian.  In
contrast, in the noncommutative deformation of ${\cal N} = 4$ SYM, which
is known to exhibit UV/IR mixing and whose nonlocality length scale
grows with the UV cutoff, we find that
entanglement entropy is extensive for all regions as long
as their size is fixed as the UV cutoff is taken away to infinity.\footnote{Entanglement 
entropy in the noncommutative theory
was studied before in \cite{Barbon:2008ut}.  Here we extend and improve on those
results.}

Recent work \cite{Lashkari:2013iga} links behaviour of entanglement
entropy to the ability of a quantum system to `scramble' information.  
Whether a given physical theory is capable of scrambling,
and how fast it can scramble, has recently became of interest to the gravity community
in the view of the so called fast scrambling conjecture
\cite{Sekino:2008he}.  It has been suggested that nonlocal
theories might emulate the scrambling behaviour of stretched black hole horizons.
While the results of \cite{Lashkari:2013iga} do not
apply directly to quantum field theories, they are quite suggestive.
Generally speaking, they imply that local (lattice) theories,
generally exhibiting area law for entanglement entropy at low energies, 
do not scramble information at these low energies, while theories with 
volume law entanglement entropy do.
As we summarized above, we demonstrate here, in the two nonlocal theories we consider,
that entanglement entropy  follows a volume law in the vacuum state.
There is no reason why entanglement entropy would cease to be extensive
in an excited energy state; if anything, high energy states are more 
likely to have extensive entanglement entropy than low-lying states
such as the vacuum state \cite{Requardt:2006ut,Alba:2009aa}.
Thus, the results of \cite{Lashkari:2013iga}
would suggest that our nonlocal theories are capable of scrambling information.
Combined with such results as those in \cite{Edalati:2012jj},
which shows that timescales for thermalization in nonlocal
theories are accelerated compared to local theories, 
our work points towards these nonlocal theories at strong coupling being fast scramblers.

Since our theories differ from ${\cal N} = 4$ SYM in the UV, 
holographic duals to we use are not asymptotically AdS spaces.
Their non-asymptotically AdS geometry has an interesting
consequence. In  previously studied examples of extensive behaviour of
entanglement entropy (for example, in thermal states) this extensive
behaviour was due to the minimal surface `wrapping' a surface in the IR region
of the dual, such as a black hole horizon (see for example \cite{Barbon:2008sr}).
Here, however, the extensivity arises from the fact that
the minimal surfaces stays close to the cutoff surface: the volume
law dependence of entanglement entropy is a UV  phenomenon.

As we were finalizing this manustript, preprint \cite{Fischler:2013gsa} appeared, which also analizes 
entanglement entropy in the noncommutative SYM and which has
some overlap with our work.

The reminder of the paper is organized as follows: in Section \ref{sec:theories}
we review nonlocal theories of interest and their gravity duals,
in Section \ref{sec:strip} we compute holographic entanglement entropy for a
strip geometry, in Section \ref{sec:cylinder} we compute holographic 
entanglement entropy in the noncommutative theory for a cylinder geometry,
in Section \ref{sec:mutualinfo} we briefly comment on mutual information in
the noncommutative theory, and in Section \ref{sec:comments} we offer
further discussion of our results.

\section{Theories considered and their gravity duals}
\label{sec:theories}

We study the strong coupling limit of two
different nonlocal deformations of
$\cal N$$=4$ SYM in 3+1 dimensions: a noncommutative
deformation and a dipole deformation.  Both of these can be realized as
the effective low energy theory on D3-branes with a 
background  NSNS B-field.  To obtain the non-commutative deformation, both
indices of the B-field must be in the worldvolume of the
D3-brane, while to obtain the dipole theory, one of the indices
must be in the worldvolume of the D3-brane while the other
one must be in an orthogonal (spacial) direction.

Since both of these theories are UV deformations of the 
$\cal N$$=4$ SYM, deep in the bulk their holographic duals
reduce to pure AdS:
\be
\frac{ds^2}{R^2} =  u^2 \left ( -dt^2 + dx^2 + dy^2 +dz^2 \right ) +
\frac {du^2}{u^2} + d\Omega_5^2~
\label{AdS}
\ee
with a constant dilaton:
\be
e^{2\phi} = g_s^2~.
\ee
In our coordinates, the boundary of  AdS space, corresponding
to UV of the field theory, is at large $u$.  It is in that region
that the holographic duals in the next two sections
will deviate from the above.

\subsection{NCSYM}

Noncommutative Yang-Mills theory is a generalization of
ordinary Yang-Mills theory to a noncommutative spacetime,
obtained by replacing the coordinates with a noncommutative algebra.
We consider a simple set up where the $x$ and $y$ coordinates
are replaced by the Heisenberg algebra, for which $[x,y] = i\theta$
and which corresponds to a noncommutative  $x-y$  plane.

One way to define this noncommutative deformation of $\cal N$$=4$ SYM
is to replace all multiplication in the Lagrangian with
a noncommutative star product:
\be
(f \star g)(x,y) ~= e^{\frac{i}{2}\theta \left(
\frac{\partial}{\partial \xi_1}
\frac{\partial}{\partial  \zeta_2} - 
\frac{\partial}{\partial \zeta_1}
\frac{\partial}{\partial  \xi_2} \right )} ~ 
f(x+\xi_1,y+\zeta_1) g(x+\xi_2,y+\zeta_2)~
|_{\xi_1 = \zeta_1 = \xi_2 = \zeta_2 = 0}
\label{star-prod}
\ee
At low energy, this corresponds to
deforming ordinary SYM theory by a gauge invariant operator of
dimension six.

The holographic dual to this noncommutative SYM  theory 
is given by the following bulk data \cite{Hashimoto:1999ut,Maldacena:1999mh}
\bear
\frac{ds^2}{R^2} &=&  u^2 \left ( -dt^2 + f(u) \left ( dx^2 + dy^2 \right )
 +dz^2 \right ) +
\frac {du^2}{u^2} + d\Omega_5^2 ~,\nn \\ \nn
e^{2\phi} &=&  g_s^2 ~f(u) ~,\\
B_{xy} &=& -\frac{1-f(u)}{\theta} = -\frac{R^2}{\alpha'}~ a_\theta^2 u^4 f(u)~,\nn\\ 
f(u) &=& \frac{1}{1 + (a_\theta u)^4}  ~,
\label{noncommutative-dual}
\eear
where $B_{xy}$ is the only nonzero component of the 
NS-NS form background. Note that 
$x,y,z$ have units of length, while $u$ has units of length inverse, or energy.
$a_\theta = (\lambda)^{1/4}\sqrt \theta$ is the weak coupling
length scale of noncommutativity $\sqrt\theta$ scaled by a power of the 't Hooft 
coupling $\lambda$ and can be thought of as the 
length scale of noncommutativity at strong coupling.

In the infrared limit, $u \ll a_\theta^{-1}$, $f(u) \approx 1$ and the
holographic dual appears to approach pure AdS space (\ref{AdS}),
while the UV region at large $u$ is strongly deformed
from pure AdS, so the holographic dual is not
asymptotically AdS.  Let $\epsilon$ denote the UV cutoff
and $u_\epsilon = \epsilon^{-1}$ the corresponding
energy cutoff.  For $\epsilon \gg a_\theta^{-1}$ ($u_\epsilon \ll a_\theta^{-1}$),
the deformed UV region of the dual spacetime is removed:
noncommutativity has been renormalized away.
However, when $u_\epsilon > a_\theta^{-1}$, 
the non-AdS geometry near the boundary can influence
the holographic computations of any field theory quantities,
including those with large length scales.  This opens the 
possibility of UV/IR mixing, defined as sensitivity of IR quantities
to the exact value of the UV cutoff.  
Noncommutative theories are known to
have UV/IR mixing \cite{Minwalla:1999px}.
The simplest way to understand the mechanism
behind the UV/IR mixing in noncommutative theories is to consider fields
with momentum $p_y$ in the $y$-direction in equation
(\ref{star-prod}): $f(x,y) = e^{-ip^f_y y} \hat f(x)$,
$g(x,y) = e^{-ip^g_y y} \hat g(x)$.  Then
$f\star g(x,y) =  e^{-i(p^g_y+p^f_y) y} \hat f(x-\theta p_y^g/2)
\hat g(x+\theta p_y^f/2)$: the interaction in the $x$-direction
is nonlocal on a length scale $\theta p_y$.
We will see that this momentum (or energy) dependence
of the scale of nonlocality is reflected in  holographic
entanglement entropy.

Finally, we need to understand the geometry of the boundary.
The metric on the boundary of the gravitational spacetime 
(\ref{noncommutative-dual}) is singular since $f \rightarrow 0$ there.
However, this is not the metric applicable to the boundary field theory, as
open string degrees of freedom see the so-called open string metric.
This is the effective metric which enters open-string correlation 
functions in the presence of a NS-NS potential B, first derived in 
\cite{Seiberg:1999vs}\footnote{For an interpretation of the open string metric
in the context of the AdS-CFT duality, see for example \cite{Li:1999am}.} and given 
by
\be
G_{ij} = g_{ij} - \left ( B g^{-1} B \right )_{ij}~,
\label{SW}
\ee
where $g_{ij}$ is the closed string metric.
Substituting our holographic data at a fixed $u$, we obtain the open
string metric, $G_{ij} = R^2 u^2 (\delta_{ij})$.  
Removing an AdS conformal factor, we see that
the boundary field theory lives on a space with a conformally invariant metric
$ds^2 = -dt^2 + dx^2 + dy^2 + dz^2$.   This is the metric we
will use to compute distances on the field theory side of the
holographic correspondence.

\subsection{Dipole theory}

Another theory we will consider is the simplest dipole deformation
of ${\cal N}=4$ SYM \cite{Chakravarty:2000qd,Bergman:2000cw,Dasgupta:2000ry}. 
 A dipole theory is one in which multiplication
has been replaced by the following non-commutative product:
\be
(f\tilde \star g)(\vec x) = f\left (\vec x- {\vec L_f \over 2} \right )
g\left (\vec x + {\vec L_g \over 2} \right)~,
\ee
where $\vec L_f$ and $\vec L_g$ are the dipole vectors assigned
to fields $f$ and $g$ respectively.
At low energy, this corresponds to a deformation by a vector
operator of dimension 5.
To retain associativity of the new product, we must
assign a dipole vector $\vec L_f + \vec L_g$ to $f\tilde \star g$.
A simple way to achieve it is to associate with each
field  $f$ a globally conserved charge $Q_f$ and to let 
$\vec L_f = \vec L Q_f$.   This can also be easily
extended to multiple globally conserved charges.
We will take $\vec L = L \hat x $ for some fixed length scale $L$,
so that our theory is nonlocal only in the x-direction.
As we saw in the previous section, noncommutative theory can
be thought of as a dipole theory with the charges being momenta
in a direction transverse to the dipole direction.\footnote{This is not entirely accurate, as
a field with transverse momentum $p$ induces a dipole moment
$\theta p$ in all the fields it interacts with instead of
in itself, but this detail will not be relevant to our
reasoning.} 

Dipole SYM is a simpler nonlocal theory than the NCSYM.
Since the scale of the noncommutativity is fixed, the theory
does not exhibit UV/IR mixing. We will see a clear signature of that in the
entanglement entropy.

The holographic dual to a dipole deformation of ${\cal N}=4$ 
SYM theory where the scalar and fermion fields
in ${\cal N}=4$ SYM are assigned dipole lengths based on
global R-symmetry charges was found, using Melvin twists,
in \cite{Bergman:2001rw}.  For the simplest case, where
supersymmetry is broken completely and where
all the scalar fields have the same dipole lengths,
the holographic dual is given by the  following bulk data:
\bear \nn
\frac{ds^2}{R^2} &=&  u^2 \left ( -dt^2 + f(u) \left ( dx^2  \right ) + dy^2
 +dz^2 \right ) +
\frac {du^2}{u^2} + \mathrm{metric~on~a~deformed~}S^5~,\\ \nn
e^{2\phi} &=&  g_s^2 ~f(u) ~,\\ 
B_{x\psi} &=& - \frac{1-f(u)} {\tilde L} = -\frac{R^2}{\alpha'} a_L u^2 f(u)~,\\ \nn
f(u) &=& \frac{1}{1 + (a_L u)^2}  ~.\nn
\eear
Similar to $a_\theta$, $a_L=\lambda^{1/2} \tilde L$, $\tilde L = L /(2\pi)$ is the 
length scale of nonlocality at strong coupling.
The usual $S_5$ of the gravity dual to a 3+1-dimensional theory
is deformed in the following way:
Express $S^5$ as $S^1$ fibration over $\mathbb{CP}^2$ (the Hopf fibration).  Then
the radius of the fiber acquires a $u$-dependent factor and is given
by $R f(u)$.  The volume of the $\mathbb{CP}^2$ is constant and given
by $R^4 \pi^2/2$.  Thus the compact manifold at radial position $u$
has a volume given by $R^5 \pi^3 f(u)$.  
$\psi$ is the global angular 1-form of the 
Hopf fibration.
For details, see 
\cite{Bergman:2001rw}.  

As we did for the noncommutative theory in the previous section,
we need to understand what metric to use for distances in
the boundary dipole theory.  Unfortunately, it does not seem
possible to give an argument similar to the one in 
\cite{Seiberg:1999vs} to find an `open string metric' for the
D-branes whose low-energy theory gives us the dipole theory,
since (in contrast to the noncommutative case) the dipole theory
requires a nonzero NSNS field $H$ and not just the nonzero potential
$B$.\footnote{A constant potential $B$ which has only one of
its indices in the worldvolume of a D-brane can be gauged away 
completely.  It is therefore important that the other index
is in a direction of a circle with varying radius, resulting
in a nonzero $H$.  In the holographic dual we consider, this
circle is the Hopf fiber.} 
The essence of the argument in \cite{Seiberg:1999vs} is that
the only effect of the potential $B$ is to change the boundary
conditions for open string worldsheet theory.  Thus, the 
boundary-boundary correlator is modified in a simple way that
is equivalent to modifying the metric.  To understand the
open string metric for the dipole set up we need a different
way to make the NSNS field $B$ `disappear'.  We can accomplish
this following  \cite{Bergman:2001rw} and  using T-duality.

First, let's see what happens when we compactify the y direction 
in (\ref{noncommutative-dual}) on a circle of radius $R_x$ and T-dualize
using the prescription in \cite{Bergshoeff:1995as}.  The T-dual
metric is
\be
 (Ru)^2 \left ( -dt^2 + (dy -(\theta/\alpha') d\tilde x)^2 +dz^2 \right ) +
\frac{1}{(Ru)^2}(d\tilde x)^2 +
\frac {du^2}{u^2} + d\Omega_5^2~,
\label{noncommutative-t-dual}
\ee
where $\tilde x$ is the T-dual coordinate to $x$.  In the
T-dual, $B$ is zero.  It has been traded for the twist 
around the $\tilde x$ circle: we identify $(\tilde x,y)$
with $(\tilde x + 2\pi \tilde R_x, y + 2\pi \tilde R_x \theta)$,
$R_x \tilde R_x \sim \alpha'$.  Conformal invariance in the $t-y-z$
directions has been restored by T-duality, and we recover the open
string metric (\ref{SW}) in those directions.\footnote{This is 
not a coincidence; the equation for the T-dual metric
\cite{Bergshoeff:1995as} and the equation for the open string
metric (\ref{SW}) are functionally the same.}  At the same time,
the twist encodes the nonlocal structure of the theory.  To see
this recall that in the noncommutative theory, fields with
momentum $p_x$ in the $x$-direction appear to have dipole
lengths $\theta p_x$.  Taking $x$ on a circle of radius  $R_x$,
$p = n/R_x$, with $n$ an integer.  When T-dualized, the corresponding
open string mode has winding number $n$ in the $\tilde x$ direction.
Given the twist, the ends of this open string are separated by 
$\Delta y = 2\pi\tilde R_x (\theta/\alpha') n$.  Substituting $n = R_x p$ we get
$\Delta y \sim  \theta p$: the twist reproduces nonlocal behaviour of
the noncommutative theory when the distances are measured in the
conformally invariant (or open string) metric.

Returning to the dipole theory, we perform T-duality in the direction
of the Hopf fiber to obtain
\be
 (Ru)^2 \left ( -dt^2 + (dx - \tilde L d\tilde \psi)^2 +dy^2 +dz^2 \right ) +
\frac{(\alpha')^2}{R^2}(d\tilde \psi)^2 +
\frac {du^2}{u^2} + d(\mathbb{CP}^2)~.
\ee
Again, the NSNS potential $B_{\psi x}$ has been replaced by a twist.
However, due to the twist of the Hopf fibration, in the T-dual there is
a new NSNS potential component, $B_{x b}$
where $b$ lies in the direction of the $\mathbb{CP}^2$,
resulting in a nontrivial NSNS field $H_{xbu}$.
Since $\psi$
was a Dirichlet direction before T-duality, the interpretation is
slightly different than it was in the noncommutative case.  After T-duality,
we have a twisted compactification identifying $(\tilde \psi, x)$ with
$(\tilde \psi + 2 \pi, x + 2\pi \tilde L)$.  
% Because $B_{xb}$ is nonzero, the twist cannot be undone without 
% introducing a new component in the NSNS potential, $B_{\tilde \psi b}$.
The proper distance between
$(\tilde \psi, x)$ and $(\tilde \psi, x+ 2\pi \tilde L)$ is therefore $\alpha'/R$,
which is small on the boundary in the large $u$ limit.  This is a sign of the
nonlocality at the dipole length $L = 2 \pi \tilde L$.  More relevant
to us at this point is that, just like for the 
noncommutative theory, conformal  invariance in the 
$t-x-y-z$ direction has been restored in the T-dual metric.
It seems reasonable then to use the metric
$-dt^2 + dx^2 +dy^2 +dz^2$ to compute distances on in the boundary theory.
For more details about this argument, as well as a string worldsheet argument
about the origin of dipole theories, see \cite{Bergman:2001rw,Dasgupta:2001zu}.

\section{Entanglement entropy for the strip}
\label{sec:strip}

We will start by studying  entanglement entropy for 
degrees of freedom living on an infinitely long strip\footnote{
In dimensions three and higher
it would be perhaps more accurate to call this region a `slab'
rather than a `strip'; nevertheless, we will use established terminology.}
$-l/2 < x < l/2 $, $-W/2 < y,z< W/2$, $W \rightarrow \infty$.
In this geometry, entanglement entropy follows the area law if
it is independent of the strip width $l$.
As we discussed in the Introduction, the relevant minimal surface is eight-dimensional; 
it wraps the compact (possibly deformed) sphere of the gravity dual
and is homologous to the strip on the boundary in the 
non-compact dimensions.  Its area is given by
\be
\mathrm{Vol}(\bar A) =  ~\pi^3 R^8 W^2 \int
_{-l/2}^{l/2} dx ~(u(x))^3 ~
\sqrt{1+\frac{(u'(x))^2}{f(u)(u(x))^4}}~,
\label{area-general-f}
\ee
where function $u(x)$ specifies the embedding of the bulk minimal area
surface.  The above formula for the area in terms of $u(x)$, with the 
appropriate form for $f(u)$, is applicable to all bulk metrics we are interested in:
while the noncommutative theory dual has more directions warped by a factor $f(u)$
than the dipole one, in the dipole theory there is another factor of $f(u)$
accounting for the deformation of the sphere on which the entangling
surface is wrapped.

Following previous work, we can think of the problem of  finding
$u(x)$ corresponding to a minimal area surface as a dynamics problem in one dimension:
$x$ plays the role of time, $u(x)$ is the position and
$u'(x)$ the velocity. Since the Lagrangian 
\be
{\cal L}(u,u') = u^3 \sqrt {1+\frac{(u')^2}{f(u) u^4}}
\ee
does not depend explicitly on the `time' $x$, there
is a conserved Hamiltonian,
\be
H = u' \frac{\partial{\cal L}(u,u')}{\partial u'} - {\cal L}(u,u') = 
- \frac{u^3}{\sqrt {1+\frac{(u')^2}{f(u) u^4}}}~.
\label{H}
\ee
The Hamiltonian $H$ is equal to $ -u_*^3$, 
where $u_*$ is the smallest value of $u(x)$ on the entangling
surface.  This point of deepest penetration of the minimal surface into
the bulk occurs at $x=0$ by symmetry.  $u'(x)$ vanishes there.

To implement the UV cutoff, differential equation (\ref{H}) is to be 
solved with a boundary condition 
\be
u(x=\pm l/2) = u_{\epsilon} = \frac{1}{\epsilon}~.
\ee

For some functions $f(u)$, equation (\ref{H}) can be integrated 
explicitly.  The answer is a hypergeometric function for $f(u) = 1$ or $f(u) = 1/(a_\theta u)^4$, 
% a Meijer G-function (for $f(u) = 1/(1+(a_L u)^2)$ or $f(u) = 1/(1+(a_\theta u)^4)$
and an elementary function for $f(u) = 1/(a_L u)^2$.
For $f(u) = 1/(1+(a_L u)^2)$ or $f(u) = 1/(1+(a_\theta u)^4)$, equation
(\ref{H}) can only be studied using series expansions in different
limits.

To compute the area of the minimal surface, it is useful to solve equation
(\ref{H}) for $u'(x)$ as a function of $u$ and substitute the result into equation
(\ref{area-general-f}).  We obtain 
\be
\mathrm{Vol}(\bar A) = 2\pi^3 R^8 W^2 \int_{u_*}^{u_\epsilon} ~\frac{du}{u'}  ~\frac{u^6} {(-H)}
= 2\pi^3 R^8 W^2 \int_{u_*}^{u_\epsilon} ~\frac{du u^4}{u_*^3} \sqrt{\frac{u_*^6}{f(u)(u^6-u_*^6)}}  ~.
\label{area-general-u}
\ee
To obtain the area of the minimal surface in terms of $l$ from this equation, given $u_\epsilon$,
it is necessary to find  the relationship between $u_*$ and $l$.

\subsection{Review of results for AdS space}

For pure AdS, with $f(u) = 1$, we can remove the boundary of AdS all the way to infinity,
$ u_{\epsilon} \rightarrow \infty$.  Then, by integrating 
(\ref{H}), we obtain a simple relationship
between $u_*$ and the width of the strip $l$:
\be
lu_* = \frac{\Gamma(2/3)\Gamma(5/6)}{\sqrt{\pi}} \approx 0.8624~.
\label{u-l-ads}
\ee
This relationship has a nice interpretation: 
holographic entanglement entropy for a structure of size $l$ is given
by the minimal surface that probes AdS bulk from the UV cutoff 
down to energy scales of order $l^{-1}$.  Modes with wavelength
longer than $l$ do not contribute to the entanglement entropy.

To compute the leading order (for $u_\epsilon \rightarrow \infty$) 
behaviour of the area of the minimal surface, we can
we can use equation (\ref{area-general-u}).  Since
$u_*$ depends only on $l$ and not on $u_\epsilon$ (i.e.,
it remains finite in the $u_\epsilon \rightarrow \infty$ limit),
the leading contribution to the area comes from large
values of $u$.  We can thus approximate 
\be
\mathrm{Vol}(\bar A) 
= 2\pi^3 R^8 W^2 \int^{u_\epsilon} ~ du u = \pi^3 R^8 ~\frac{W^2 }{ \epsilon^2}~.
\label{area-ads}
\ee
A more precise result for the entanglement entropy density 
is obtained from a next-to-leading order computation.
It gives a universal term which is finite and independent of the cutoff:
\be
\frac{S}{W^2} = \frac{R^3 }{4G_N^{(5)}}  \left [ 
\frac{1}{\epsilon^2} ~-~  \frac{
\left(2\Gamma\left(2\over 3\right)\Gamma\left(5 \over 6\right)\right)^3 }
{\pi^{3/2}} \frac{1}{l^2}  ~+~ (\mathrm {terms~that~
vanish~for~}\epsilon \rightarrow 0)
 \right ]~.
\ee
In terms of field theory variables, we have 
\be
 \frac{R^3 }{4G_N^{(5)}} = \frac{N^2}{2\pi}~,
\label{newton-N}
\ee
so that the divergent part of the entanglement entropy is
proportional to $N^2 \epsilon^{-2}$, with a numerical coefficient
which is specific to strongly coupled ${\cal N} = 4$ SYM.  The
entanglement entropy is therefore of this generic form (applicable to
3+1 dimensions):
\be
S =  N_{\mathrm{eff}} \frac{|\partial A|}{\epsilon^2} = 
N_{\mathrm{eff}} \frac{W^2}{\epsilon^2}
\label{S-generic}
\ee
with the number of effective on-shell degrees of freedom
$N_\mathrm{eff}$ proportional to $N^2$.

Formula (\ref{S-generic}) supports the following simple picture
of entanglement entropy in theory with a local UV fixed point:
A quantum field theory in 3+1 dimensions with a UV cutoff 
$\epsilon^{-1}$ can be thought of as having on the order of
one degree of freedom per cell of volume $\epsilon^{3}$.
The divergent part of the geometric entanglement entropy 
computed the vacuum state of such a theory  is a measure 
of the effective number of degrees of freedom inside
of a region $A$ that have quantum correlations with 
degrees of freedom outside of $A$.
In a local theory, only `adjacent' degrees of freedom
are coupled via the Hamiltonian and the simple
intuition is that therefore quantum correlations between
degrees of freedom inside of $A$ and outside of $A$ happen only
across the boundary $\partial A$.  Thus, the dominant part 
of the entanglement entropy 
comes from degrees of freedom which live within a distance
$\epsilon$ of the boundary of $A$, with entanglement entropy
proportional to the volume of this `skin' region, 
equal to $\epsilon |\partial A|$.  Dividing this
volume by the volume of one cell, $\epsilon^3$, gives equation
(\ref{S-generic}).

\subsection{Dipole theory}

Having briefly reviewed holographic entanglement entropy on a strip in undeformed SYM,
we will now study it in the dipole theory.

\begin{figure}
\center{\includegraphics[scale=0.4]{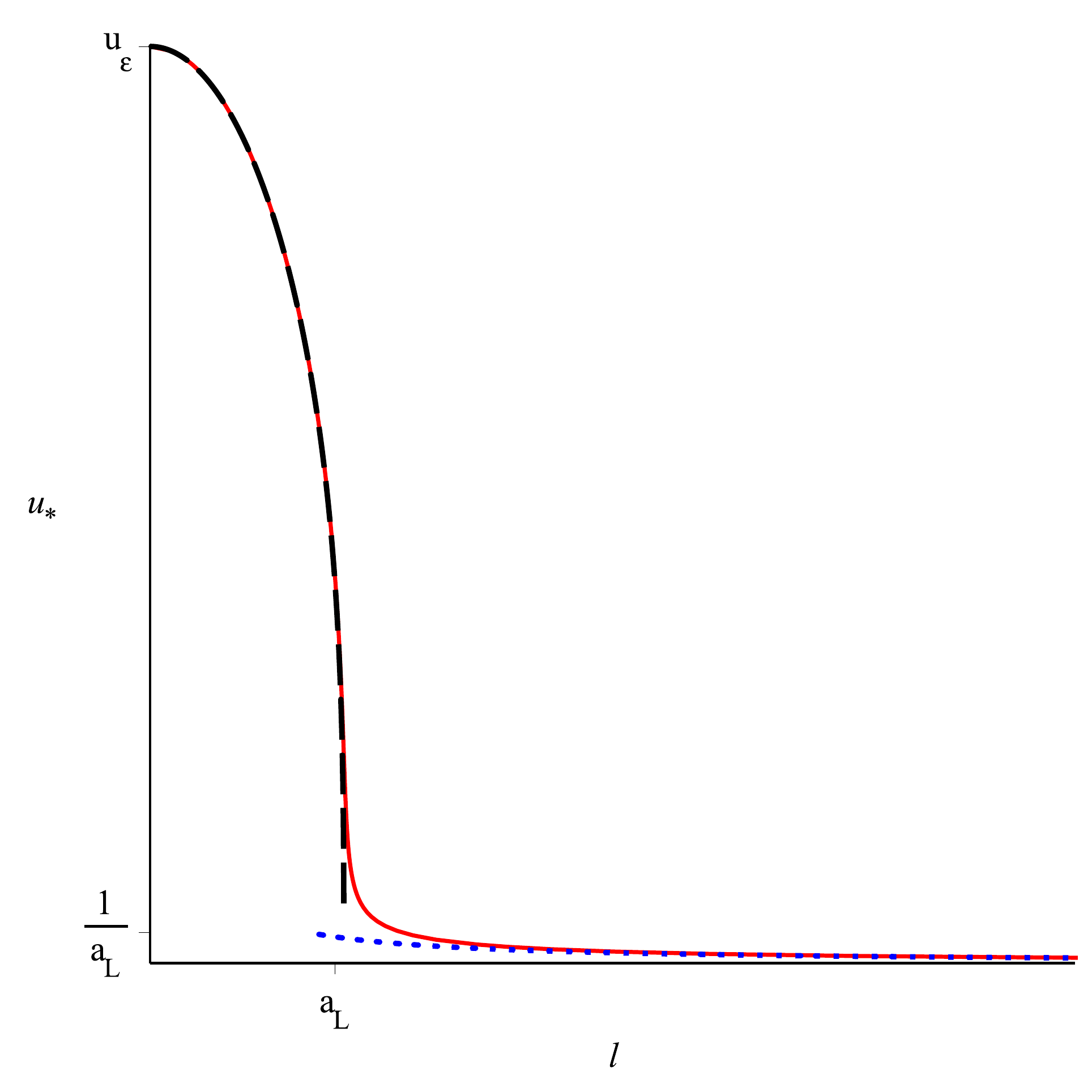}}
\caption{Point of deepest penetration $u_*$ as a function of the strip width $l$ for
a minimal area surface in the gravity dual to the dipole theory (solid red line).  The blue
dotted line shows the result for pure AdS, given by equation (\ref{u-l-ads}), while
the black dashed line shows the narrow strip approximation, equation (\ref{u-l-dipole-strip-narrow}).
In this Figure, $a_L u_\epsilon = 30$.
}
\label{f1}
\end{figure}

In Figure \ref{f1}, we show the relationship between
$l$ and $u_*$ for the dipole theory.  We see that it
approaches the AdS result (\ref{u-l-ads}) for large strip widths $l$ and that
it deviates significantly from it for strips whose width is on the order of and
smaller than $a_L$.  For  narrow strips, the entangling
surface does not penetrate the bulk very deeply into the IR
region.  To study these, we assume that $u_* \gg a_L^{-1}$ and
write $f(u) \approx (a_L u)^{-2}$.  Here we get a pleasant
surprise: the exact shape of the minimal surface
can be obtained in terms of elementary functions
\be
u(x) = \frac{u_*}{\cos(3 x/a_L)^{1/3}}~~\mathrm{for}~x/a_L \in [-\pi / 6,\pi/6]~.
\ee
The relationship between the penetration
depth of the minimal surface and the width of the strip is
\be
u_* = u_{\epsilon} \left (\cos(3 l / 2 a_L) \right )^{1/3}~.
\label{u-l-dipole-strip-narrow}
\ee
This equation is valid as long as $u_* \gg a_L^{-1}$, which, in the
limit where $u_{\epsilon}$ is large,  is true 
for all strip widths $l$ up to $l = (\pi / 3) a_L$.   Notice that,
in contrast to pure AdS, the point of deepest penetration $u_*$ depends on
the UV cutoff.  Thus, if one works
at the limit of infinite cutoff, these minimal area surfaces will be missed.

The area of the minimal surface under the approximation $f(u)\approx (a_L u)^{-2}$ is
\be
\mathrm{Vol}(\bar A) =  \pi^3 R^8  \left [ \frac{ W^2 a_L}{\epsilon^3}~
\frac{2\sin(3l/2a_L) }{3}
\right ] \approx
\pi^3 R^8  ~  \frac{ W^2l}{\epsilon^3}~,
\label{A-dipole-strip-narrow}
\ee
where the final approximation is for a small strip width $l \ll a_L$.
For narrow strips, entanglement entropy is extensive, proportional
to the width of the strip.  The first part of equation 
(\ref{A-dipole-strip-narrow}) gives the corrections to the volume
scaling, controlled by the powers of the ratio $l/a_L$.

For surfaces with large l (compared to $a_L$), we can use the same 
approximation as in equation (\ref{area-ads}), with $f(u) \approx (a_L u)^{-2}$:
\be
\mathrm{Vol}(\bar A) 
= 2\pi^3 R^8 W^2 a_L \int^{u_\epsilon} ~ du u^2 = \pi^3 R^8 ~\frac{2 W^2 a_L}{3 \epsilon^3}
\label{A-dipole-strip-wide}
\ee

\begin{figure}
\center{\includegraphics[scale=0.4]{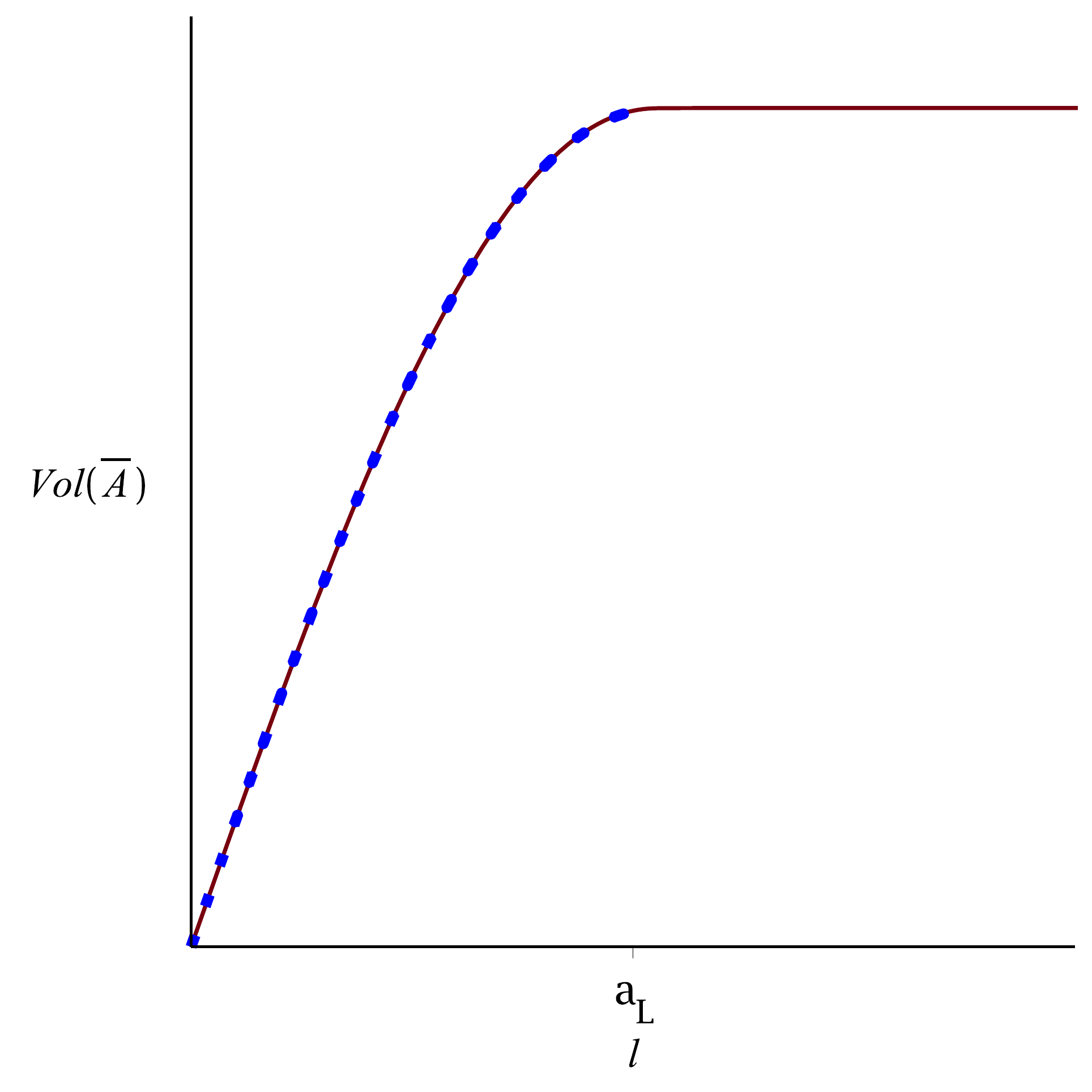}}
\caption{The area of the minimal surface as a function of the strip width $l$ for
for the dipole theory (solid red line).  The blue
dotted line shows result (\ref{A-dipole-strip-narrow}), 
valid for narrow strips $l<(\pi/3)a_L$. In this Figure, $a_L u_\epsilon = 30$.
}
\label{f2}
\end{figure}

We see that this area, which is independent of the width, is the same as the area obtained
from equation (\ref{A-dipole-strip-narrow}) at the extremal value of $l$, $l=a_L \pi/3$.  
The situation is illustrated in Figure \ref{f2}.

To summarize, we obtained the following result for the entanglement
entropy density in the  dipole theory:
\be
\frac{S}{W^2} = \frac{N^2}{2\pi} ~\frac{2a_L}{3 \epsilon^3} ~G(l/a_L)~,
~~\mathrm{where}~G(z) = \left \{
\begin{array}{ll}
\sin(3z/2) & ~\mathrm{for}~z<\pi/3~,\\
1 & ~\mathrm{for}~z>\pi/3~.
\end{array}
 \right .
\label{S-dipole}
\ee
Entanglement entropy is extensive for very narrow strips, 
depends on the width of the strip
in a nonlinear fashion for widths up to the nonlocality scale and smoothly
goes over to a non-extensive (area law) behaviour for wide strips.
For wide strips, while the entanglement entropy follows an
area law, it has a different form than it would for a 
a generic local theory (given by equation (\ref{S-generic})).
To explain this, apply reasoning similar to that below equation (\ref{S-generic})
to a theory with a fixed scale of nonlocality $a_L$.  By definition, the
Hamiltonian of such a theory couples together degrees of freedom
as far apart at $a_L$, thus, for a large region,
the dominant part of geometric entanglement 
entropy should be proportional to the volume of a set of points 
no more than $a_L$ away from the boundary of $A$.  This volume,
for a large enough region, can be approximated by $a_L|\partial A|$,
leading to $S \propto a_L|\partial A| / \epsilon^3$, which is
exactly what we see in equation (\ref{S-dipole}) for a strip with $l > (\pi/3) a_L$.

Applying our reasoning to the narrow strip, we see that,
for $l<a_L$, all degrees of freedom inside the 
strip should are directly interacting with, and therefore
entangled with, degrees of freedom outside of the strip.
For a very narrow strip, degrees of freedom inside it
will mostly be entangled with the degrees of freedom outside,
and entanglement entropy should be proportional to $l$, which is
exactly what we see.  As the strip gets wider, some of the 
degrees of freedom inside the strip become entangled with each other,
decreasing the entanglement with the outside and implying a
sub-linear growth to the entanglement entropy as a function of 
$l$, again in agreement with equation (\ref{S-dipole}).  

The exact way in which $S$ deviates from $S \propto l$ 
can be viewed as a way to probe the distribution of quantum correlations
in the ground state of this nonlocal theory.
It would be interesting to consider this further.

Finally, notice that above the nonlocality length scale $a_L$, the
shape of the minimal surface is not greatly affected
by the exact value of the cutoff; this is a sign that
the dipole theory does not have UV/IR mixing.  We will
see a very different behaviour  for the noncommutative theory.

\subsection{NCSYM}

\begin{figure}
\center{\includegraphics[scale=0.45]{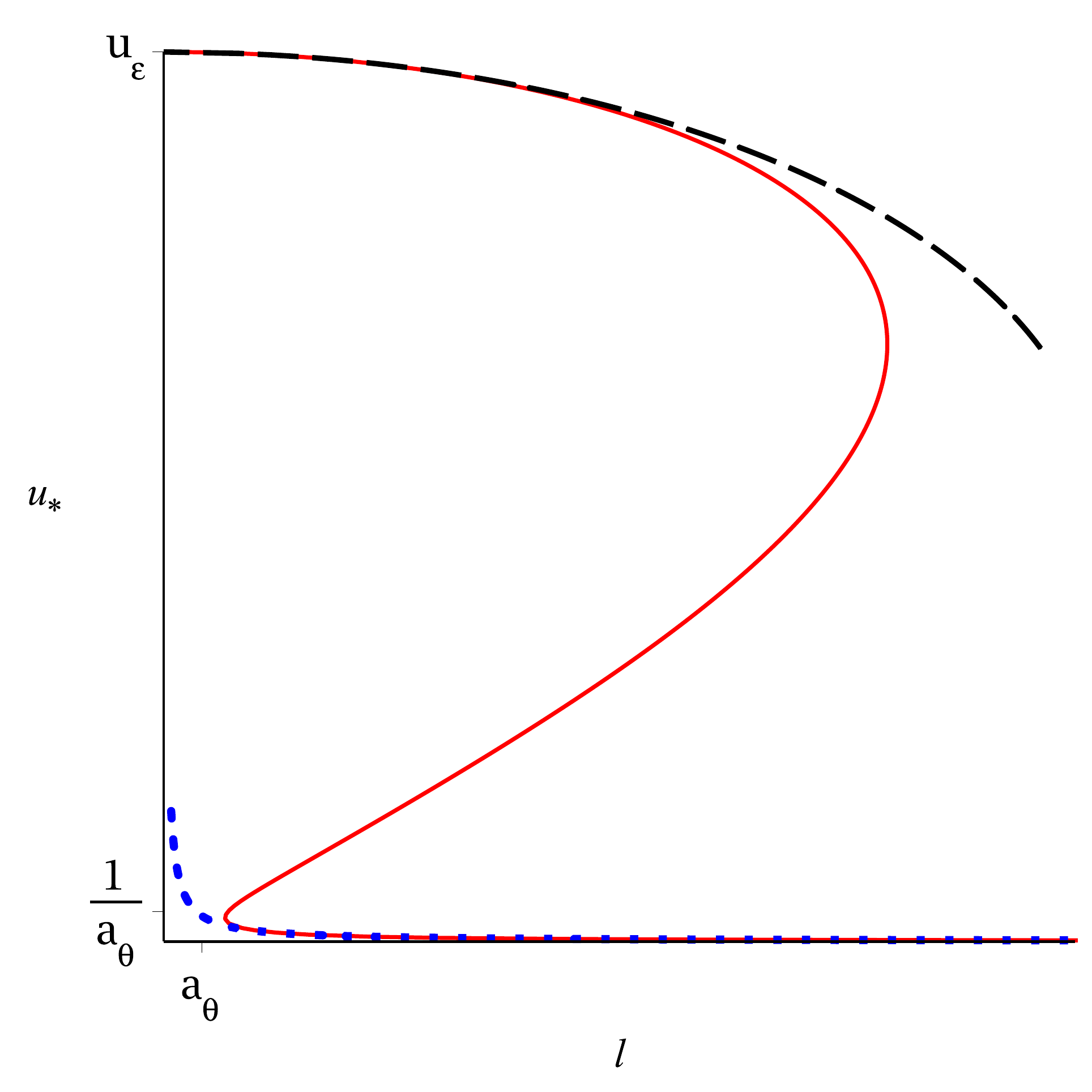}}
\caption{Point of deepest penetration $u_*$ as a function of the strip width $l$ for
extremal area surfaces in the gravity dual to the noncommutative theory (solid red line).  The blue
dotted line shows the result for pure AdS, given by equation (\ref{u-l-ads}), while
the black dashed line shows the result of equation (\ref{u-l-ncym}).
In this Figure, $a_\theta u_\epsilon = 30$.
}
\label{f3}
\end{figure}

\begin{figure}
\center{\includegraphics[scale=0.4]{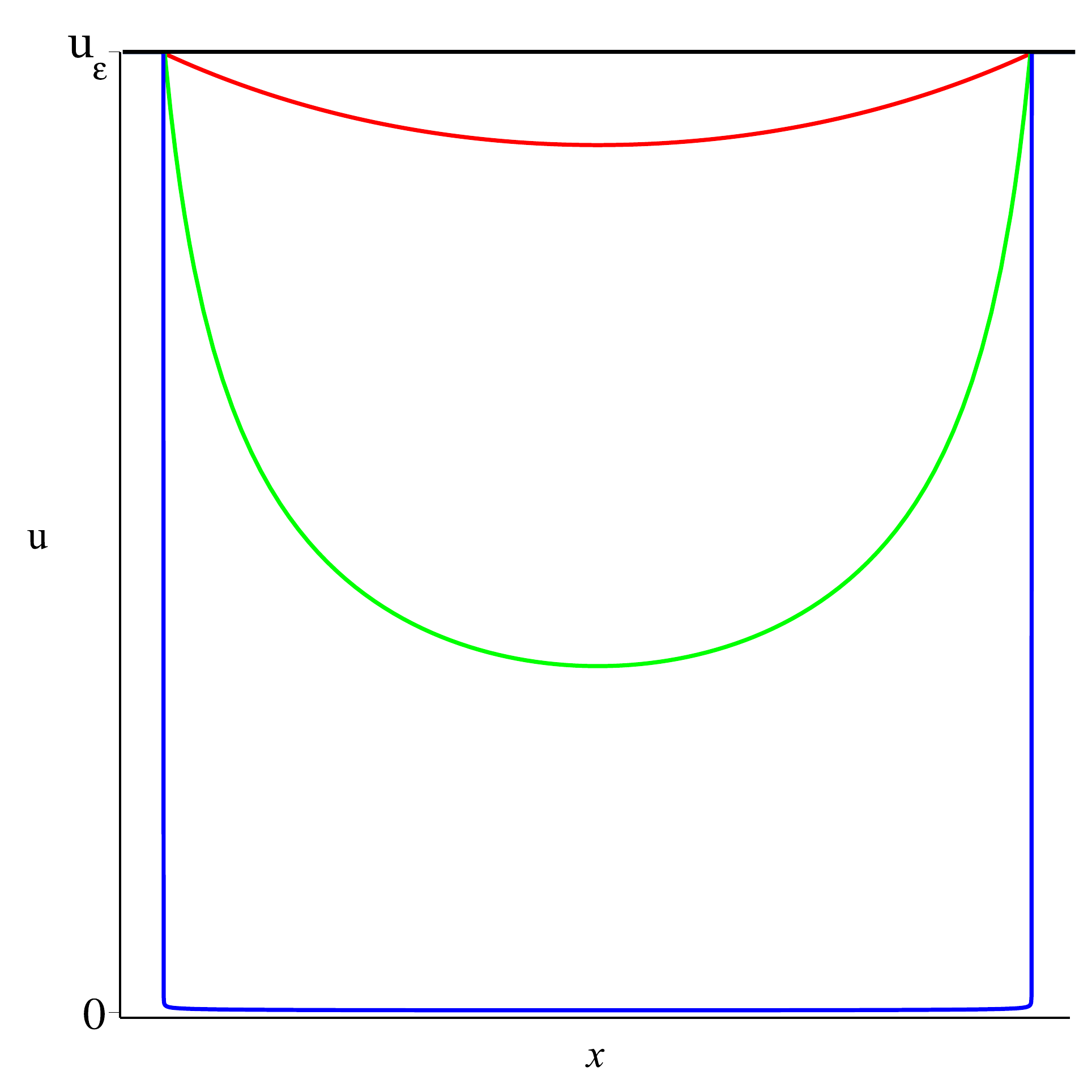}}
\caption{Shape of three extremal area surfaces, given
as $u(x)$, all anchored on the same boundary strip.  
}
\label{f4}
\end{figure}

For entanglement entropy of a strip in the noncommutative theory, 
the situation is more complicated.  As is shown in
Figure \ref{f3}, there are as many as three extremal area surfaces for a given
width $l$ of the strip. 
At large strip widths there is only one surface, for which the relationship
between $l$ and $u_*$ approaches that of pure AdS, given by equation (\ref{u-l-ads}).
At small widths, similarly to the dipole theory, there is a surface
which stays close to the cutoff surface.\footnote{In \cite{Barbon:2008ut}, this surface
was approximated by one placed exactly at the cutoff, at constant $u$.}  
At intermediate $l$, there are three extremal surfaces, whose shape is
shown in Figure \ref{f4}.
As we will see, the middle of the three surfaces is always unphysical
(its area is never smaller than the other two).  As the width is increased
from zero, at some critical width $l_c$ there is a phase transition as the area
of the surface on the top-most branch becomes larger than the area of the 
surface on the bottom-most branch in Figure \ref{f3}.

We start by studying top-most branch, which contains
surfaces anchored on narrow strips.  To study these,
we find $u(x)$ as a series expansion for small $x$.  This allows us to write the
relationship between $l$, $u_*$  and  $u_\epsilon$ for small $l$:
\be
u_\epsilon - u_* = \frac{3}{8} \frac{u_*^3}{1+(a_\theta u_*)^4} l^2 + {\cal O}((l/a_\theta)^4)~.
\label{u-l-ncym}
\ee
The integral in equation (\ref{area-general-f}) can also be expanded and evaluated for
small $l$.  Finally, substituting $u_*$ from the expression above into
the area integral, we can obtain the area for small $l$:
\be
\mathrm{Vol}(\bar A) =  ~\pi^3 R^8 W^2 \left [
\frac{l}{\epsilon^3} ~-~ \frac{3}{8} \frac{l^3}{a_\theta^4\epsilon(1 + (\epsilon/a_\theta)^4)} 
+ {\cal O}((l/a_\theta)^4) \right ]~.
\label{area-ncym-l-expansion}
\ee
We have kept the sub-leading terms in $\epsilon$ for completeness---expression (\ref{area-ncym-l-expansion}),
as given, is correct even for large $\epsilon$ as long as $l$ is small.  

From equation (\ref{u-l-ncym}) we see that as we increase $u_\epsilon$ keeping $l$
fixed, $(u_\epsilon - u_*) \propto  l^2 / u_*$, so that $u_\epsilon - u_*$ approaches zero:
the minimal surface approaches the boundary surface.

This result turns out to hold even for large (but fixed) strip width $l$ in 
the large $u_\epsilon$  limit.  In this limit, we approximate $f(u) \approx (a_\theta u)^4$.
This allows us to obtain $l$ and the area as a function of $u_*$ and $u_\epsilon$
in terms of hypergeometric functions.  We see that $l/u_\epsilon$ is a function of the
ratio $u_*/u_\epsilon$ only.  As $u_\epsilon$ approaches infinity with $l$ fixed,
this ratio goes to $1$, showing that the entire minimal surface stays close
to the boundary and that our approximation $f(u) \approx (a_\theta u)^4$ is self-consistent
even for large $l$, as long as $l$ is held fixed when the 
UV cutoff is removed.  The following relationship holds under this approximation:
\be
\int_{-l/2}^{l/2} dx ~{\cal L}  = \frac{l u_*^3 + u_\epsilon\sqrt{u_\epsilon^6-u_*^6}}{4} ~.
\ee
Thus, the leading order UV divergence of the area of the minimal surface 
at any fixed width $l$ is
\be
\mathrm{Vol}(\bar A) =  ~\pi^3 R^8 W^2 \frac{l}{\epsilon^3}~.
\label{A-ncym-strip-shallow}
\ee

Having understood the top-most branch of the plot in Figure \ref{f3} , corresponding 
to surfaces that stay close to the boundary, we now move to
the bottom-most one.  These surfaces penetrate deeply into the bulk and their
shape is not affected by the cutoff point.  We can therefore use the 
same method as before for obtaining their area:
\be
\mathrm{Vol}(\bar A) 
= 2\pi^3 R^8 W^2 a_\theta^2 \int^{u_\epsilon} ~ u^3 du  = \pi^3 R^8 ~\frac{ W^2 a_\theta^2}{2 \epsilon^4}~.
\label{A-ncym-strip-deep}
\ee
\begin{figure}
\center{
\includegraphics[scale=0.45]{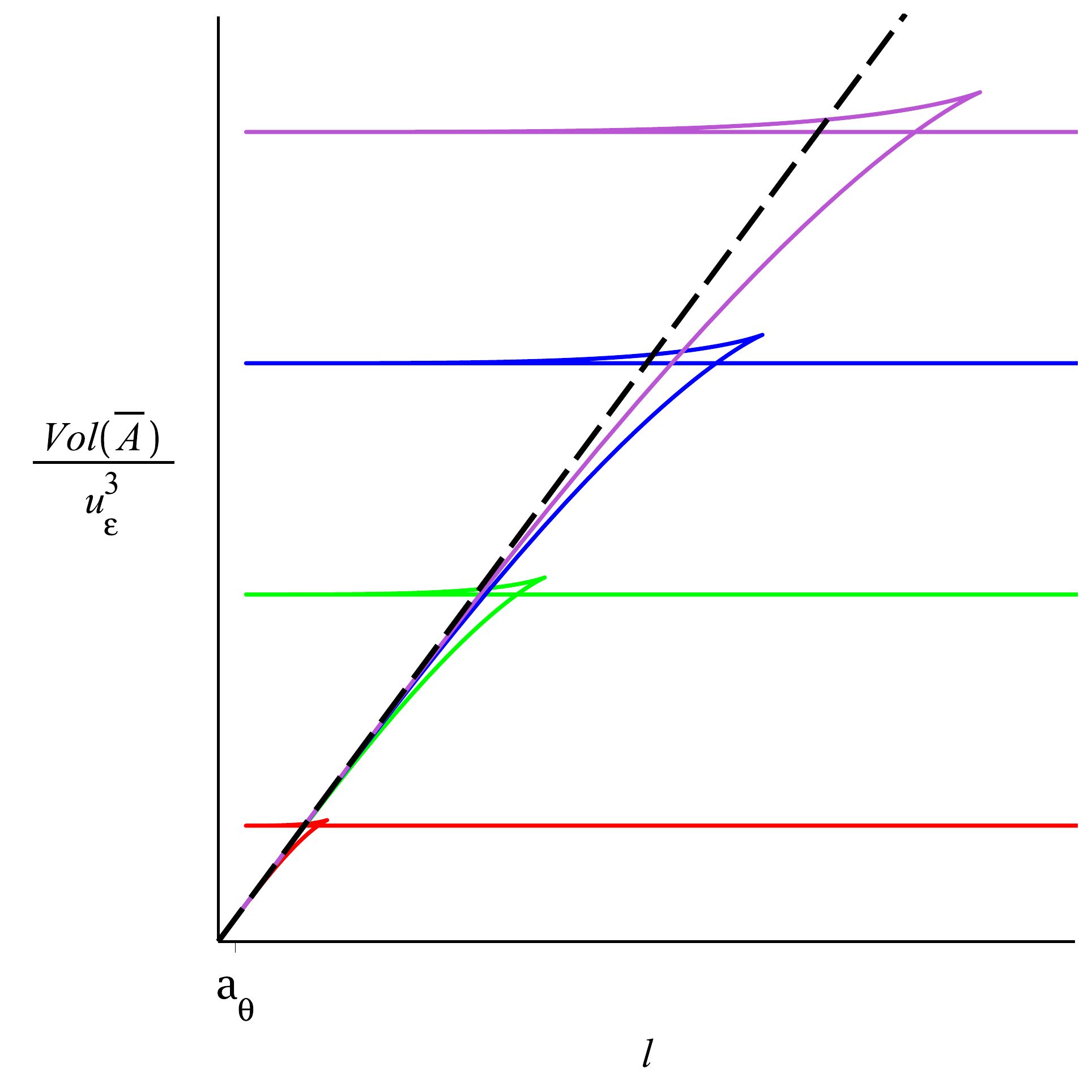}

\includegraphics[scale=0.45]{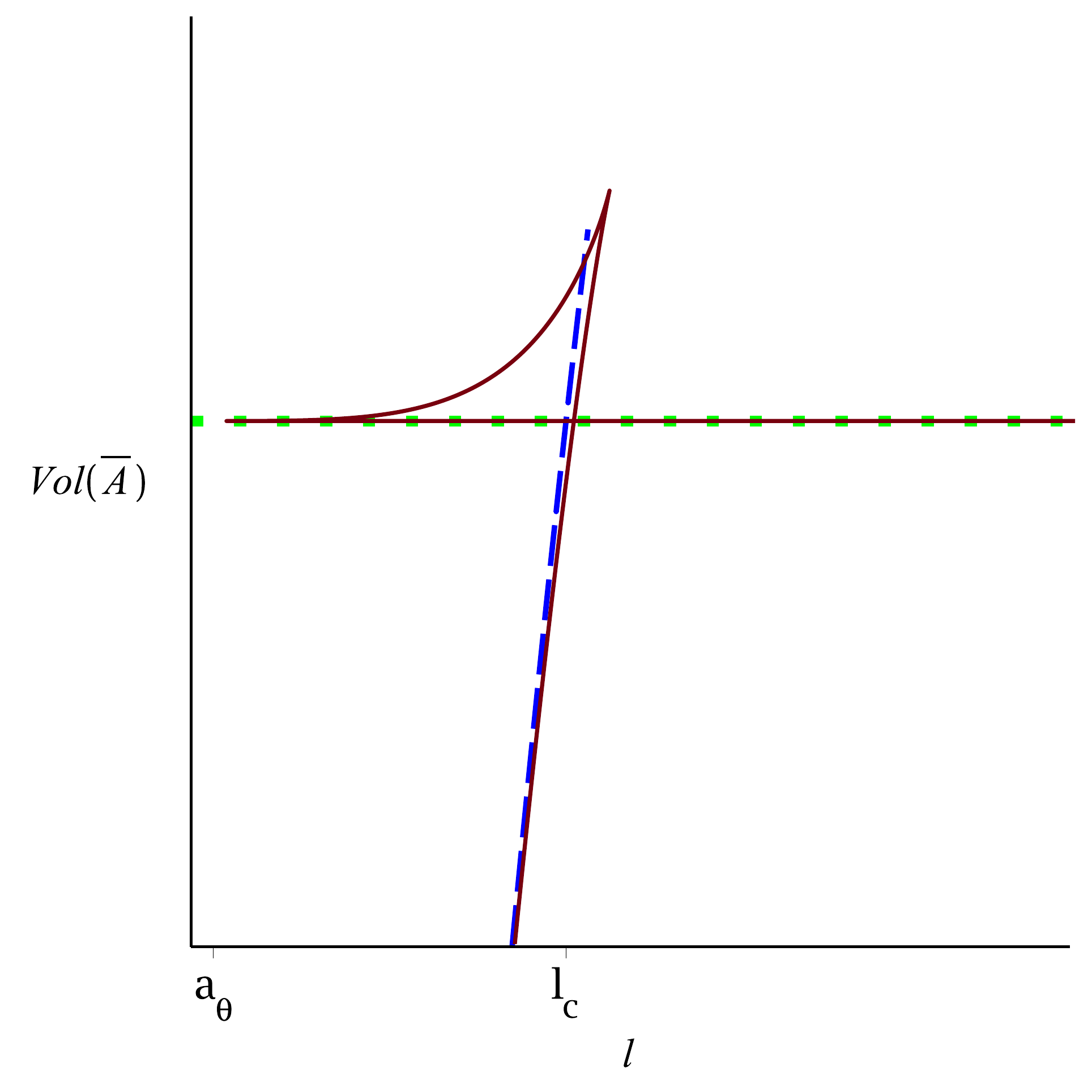}}
\caption{Area of the minimal surface as a function of strip width $l$
for noncommutative theory.  
Top: Plots with $a_\theta u_\epsilon = 10$, $30$, $50$ and
$70$ are shown.  Area is scaled by a power of the cutoff to 
allow functions for different cutoffs to be plotted on the
same set of axis.  Dashed line corresponds to the leading
term in equation (\ref{area-ncym-l-expansion}), 
$\mathrm{Vol}(\bar A)/u_\epsilon^3 \propto l$.
The range of validity of this approximate expression increases
with increasing $u_\epsilon$. Bottom: Detail of the fish-tail
phase transition is shown. The green dotted line and the blue dashed line correspond to
equations (\ref{A-ncym-strip-deep}) and (\ref{area-ncym-l-expansion}) respectively.
$a_\theta u_\epsilon = 30$.}
\label{f5}
\end{figure}

Since there are  multiple extremal surfaces anchored on a strip,
we need to find out which of them have the smallest area at a given $l$.
At very small $l$ there is only one surface (see Figure \ref{f3}), 
thus, by continuity,  for $l$ less than some critical length $l_c$,
the surface of the smallest area corresponds to the top-most branch of 
Figure \ref{f3}.  Its area is given by equation (\ref{A-ncym-strip-shallow}).
At $l_c$ there is a first order phase transition.\footnote{This is similar to
\cite{Narayan:2012ks} and to \cite{Klebanov:2007ws}, as well as to
the recent paper \cite{Hubeny:2013gta}.  Entanglement entropy is
continuous across the transition, but its derivative is not.}
Above $l_c$,  the surface with
the smallest area is on the bottom-most branch of Figure \ref{f3} and
its area is given by equation (\ref{A-ncym-strip-deep}).  To compute $l_c$,
we set  equations (\ref{A-ncym-strip-shallow}) and (\ref{A-ncym-strip-deep})
equal and obtain that $l_c = a_\theta^2 u_\epsilon / 2$. 

Since the critical length increases with $u_\epsilon$, if we hold $l$ fixed and
take the limit $u_\epsilon \rightarrow \infty$, $l_c$ will diverge to 
infinity as well and equation (\ref{A-ncym-strip-shallow}) will hold
for any $l$.

Our analysis implies that in the limit $\epsilon \rightarrow 0$, the
entanglement entropy density for a strip of a fixed length $l$ is
\be
\frac{S }{W^2} = \frac{N}{2\pi}~\left [ \frac{l}{\epsilon^3}~-
\frac{3}{8}\frac{l^3}{a_\theta^4 \epsilon} ~+~
\mathrm{terms~vanishing~for~}\epsilon\rightarrow 0 \right ]~,
\label{S-NCYM}
\ee
which, to the leading order, is the same answer as for the dipole theory in the narrow strip
limit (equation (\ref{S-dipole}), $l \ll a_L$).

To understand the physics behind this result, we recall that in the noncommutative theory
a mode with momentum $p_y$ in the $y$-direction can be thought of as
a dipole field with an effective dipole length $\theta p_y$
in the $x$-direction. 
The high-momentum modes which dominate the 
divergent part of entanglement entropy all have large effective dipole
moments.  Therefore the entanglement entropy is that of a nonlocal 
theory with a large effective scale of nonlocality.  This is precisely 
what we see.

In the complementary limit, fixing a (large) UV cutoff first and then considering
wide strips, $l>l_c$,  equation (\ref{A-ncym-strip-deep}) shows that 
entanglement entropy density is equal to 
\be
\frac{S}{W^2}  = \frac{N^2}{2\pi} ~\frac{a_\theta^2}{2 \epsilon^4}~.
\label{S-NCYM-wide-strip}
\ee
We see that the area law applies and the number of degrees of freedom
which are near enough to the boundary of the region to be entangled
with the outside is proportional to $a_\theta^2/\epsilon^2$.  This is 
equal to the scale of noncommutativity at the UV cutoff ($a_\theta^2 u_\epsilon$=
$a_\theta^2/\epsilon$) divided by the cutoff length scale $\epsilon$, 
consistent with our previous discussions.

In the next section, we will compute the entanglement entropy in 
the noncommutative theory for another geometry: a cylinder whose
circular cross-section is in the two noncommutative directions
$x$ and $y$ and which is extended infinitely in the commutative direction $z$.
We will obtain a result for the entanglement entropy that is similar
to the one in this section, while the geometry of the entangling
surfaces will be very different.

\section{Entanglement entropy for the cylinder in NCSYM}
\label{sec:cylinder}

Consider a region on the boundary extended in the $z$ direction 
($-W/2<z<W/2$, $W\rightarrow \infty$) and 
satisfying $x^2 + y^2 < l^2$ in the $x$ and $y$ directions.  The
area functional for a surface homologous to this cylindrical region,
assuming rotational symmetry in the $x-y$ plane and translational
symmetry in the $z$ direction, is
\be
\mathrm{Vol}(\bar A) = 2\pi^4 R^8 W \int_{0}^l dr ~r~(u(r))^3 ~
\sqrt{1+\frac{(u'(r))^2}{f(u)(u(r))^4}}~,
\ee
where $r = \sqrt{x^2+y^2}$ and the surface is specified by
 a function $u(r)$.

Since $r$ appears explicitly in the Lagrangian density, the 
equation of motion cannot be integrated explicitly even once.
We will therefore have to rely more on numerical computation.

\begin{figure}
\center{\includegraphics[scale=0.4]{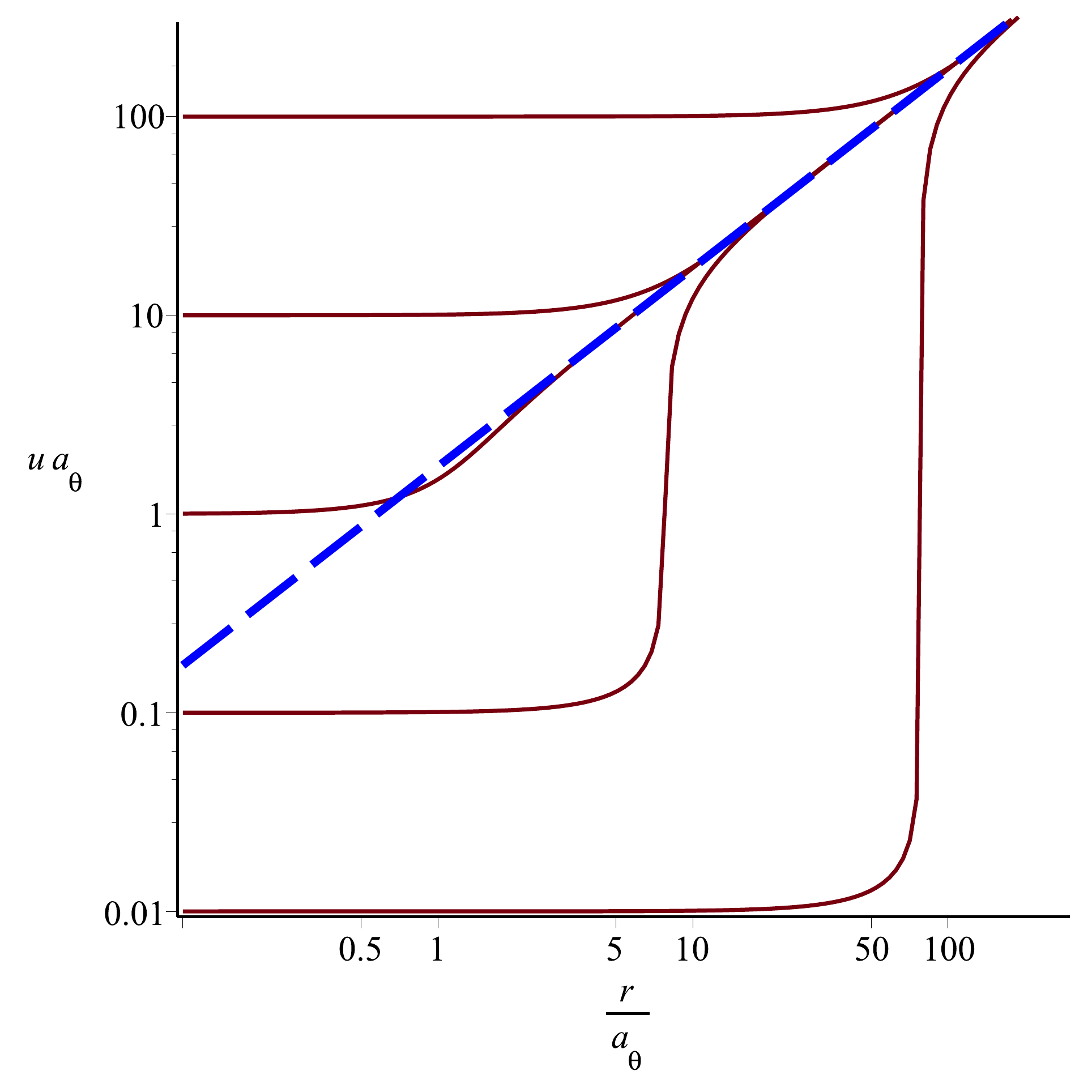}
\includegraphics[scale=0.4]{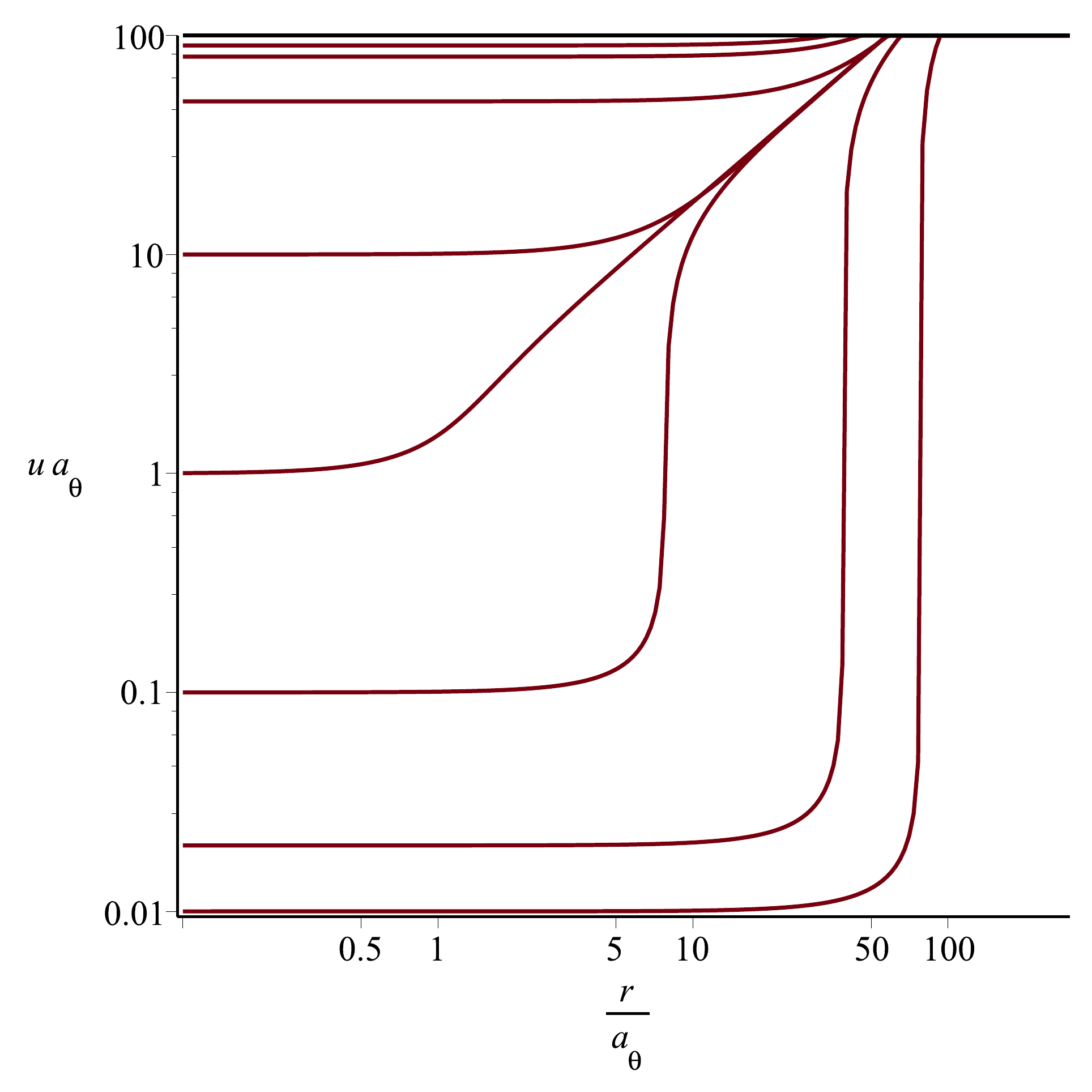}}
\caption{Extremal surfaces homologous to a cylinder in NCSYM, presented as $u(r)$.
On the left, the straight dashed line is the asymptotic behaviour 
given by $a_\theta u=\sqrt 3 r/a_\theta$.
On the right, surfaces with $l$ sufficiently smaller or
larger than $l_c = a_\theta^2 u_\epsilon / \sqrt{3}$ 
to reach the cutoff before they had time to approach the this
asymptote are shown as well.
}
\label{f7}
\end{figure}

Figure \ref{f7} shows shapes of extremal surfaces anchored
on a disk in the boundary noncommutative theory.  As is easy to check
analytically, all these surfaces asymptote at large $r$ and $u$ 
to a single 'cone' given by $a_\theta u = \sqrt 3 r/a_\theta$.  A linear analysis
about this asymptotic solution gives
\be
a_\theta u(r) \approx \sqrt 3 r/a_\theta + t  \cos \left ( \frac{\sqrt 7}{2} \ln (r/a_\theta) 
+  \varphi \right ) ~,
\label{oscillations}
\ee
where $t$ and $\varphi$ are free parameters, with $t$ small.  In principle,
a relationship between $t$ and $\varphi$ could be derived given that
$u'(0) = 0$, but it cannot be obtained
within perturbation theory.  It is interesting and perhaps surprising
that the fluctuations about the asymptote are oscillatory in $r$.
This behaviour, which can be seen in Figure \ref{f7}, 
agrees very well with more detailed numerical analysis.

From Figure \ref{f7} we see that surfaces with $u_*$ relatively close to $a_\theta^{-1}$
approach the asymptote $u = \sqrt{3} r / a_\theta^2$ before reaching the cutoff, while
those with large $u_*$ ($a_\theta u_* \gg 1$) or small $u_*$ ($a_\theta u_* \ll 1$)
do not.   At a fixed cutoff, then, we have three classes of surfaces: 
shallowly probing surfaces, $a_\theta u_* \gg 1$, with $l$ smaller
than and bounded away from $l_c := a_\theta^2 u_\epsilon / \sqrt{3}$,
deeply probing surfaces, $a_\theta u_* \ll 1$, with  $l$ larger than than and bounded away from $l_c$ 
and the intermediate category, for which $l$ is approximately equal to $l_c$.
In the first and second category, there is a unique extremal surface
at a given radius $l$, while for radii close to $l_c$ the situation
is more complicated, due to the oscillatory nature of the 
near-asymptotic solutions shown in equation (\ref{oscillations}).  
Since the cutoff radius  $l_c$ increases with $u_\epsilon$ (similar to
the behaviour in the strip geometry), the entanglement surface for a region with
any radius $l$ belongs to the first category for a sufficiently high 
UV cutoff.  

First, let us consider the surfaces with small $l/a_\theta$.  These
can be studied by expanding in  $l/a_\theta$.  We get the 
following two results:
\be
u_\epsilon - u_* = \frac{3}{4} \frac{u_*^3}{1+(a_\theta u_*)^4} l^2 + {\cal O}((l/a_\theta)^4)~,
\label{u-l-expansion-circle}
\ee
\be
\mathrm{Vol}(\bar A) =  ~2\pi^4 R^8 W \left [
\frac{l^2}{2\epsilon^3} ~-~ \frac{9}{32} \frac{l^4}{a_\theta^4\epsilon(1 + (\epsilon/a_\theta)^4)} 
+ {\cal O}((l/a_\theta)^4) \right ]~.
\label{area-l-expansion-circle}
\ee
The $l/a_\theta$ expansion for the area of the minimal surface has a structure which
is similar to the one we obtained for the strip in the noncommutative theory: 
organizing the expansion in powers of $l$, the term of order $l^n$ has
as its leading $\epsilon$ dependence $1/\epsilon^{5-n}$
(with $n$ even).  Assuming that this analytic structure is valid for finite
$l/a_\theta$, we obtain that in the limit $\epsilon \rightarrow 0$, the
entanglement entropy density for a cylinder at a fixed radius $l$ is
\be
\frac{S }{W} = \frac{N}{2\pi}~\left [ \frac{\pi l^2}{\epsilon^3}~-~
\frac{9}{32}\frac{l^4}{a_\theta^4 \epsilon} +
\mathrm{terms~vanishing~for~}\epsilon\rightarrow 0 \right ]~.
\label{S-NCYM-circle}
\ee
Qualitatively, this is the same answer as we obtained for the strip: entanglement
entropy is extensive, proportional to the volume of the considered region.
Notice that neither expression has a non-zero universal (independent of $\epsilon$ part).

At finite (and large) cutoff, we can  consider large radius cylinders.
For $l$ sufficiently larger than $l_c$ we see from Figure \ref{f7} 
that $u_* a_\theta \ll 1$ and the entangling 
surface seems close in shape to that in pure AdS (as it approaches the
boundary at approximately the right angle, based on numerical evidence).  Thus, 
$u_* \propto l^{-1}$ and the area is approximately
\be
\mathrm{Vol}(\bar A) =  ~2\pi^4 R^8 W a_\theta^{2} \int dr ~r~u^3~ u'(r) = 
2\pi^4 R^8 W a_\theta^{2} l \int ^{u_\epsilon} ~ du u^3  = \pi^3 R^8 ~
\frac{2 \pi l W  a_\theta^2 }{4  \epsilon^4}~,
\label{area-l-large-circle}
\ee
where we have used $f(u) \approx (a_\theta u)^{-4}$ and approximated $r \approx l$ in the 
region near the boundary.  Resulting entanglement entropy has the same interpretation
as the one in equation (\ref{S-NCYM-wide-strip}), with the area of the strip's boundary, 
$W^2$ replaced by the area of the boundary of the cylinder, $2\pi l W$.

\begin{figure}
\center{\includegraphics[scale=0.4]{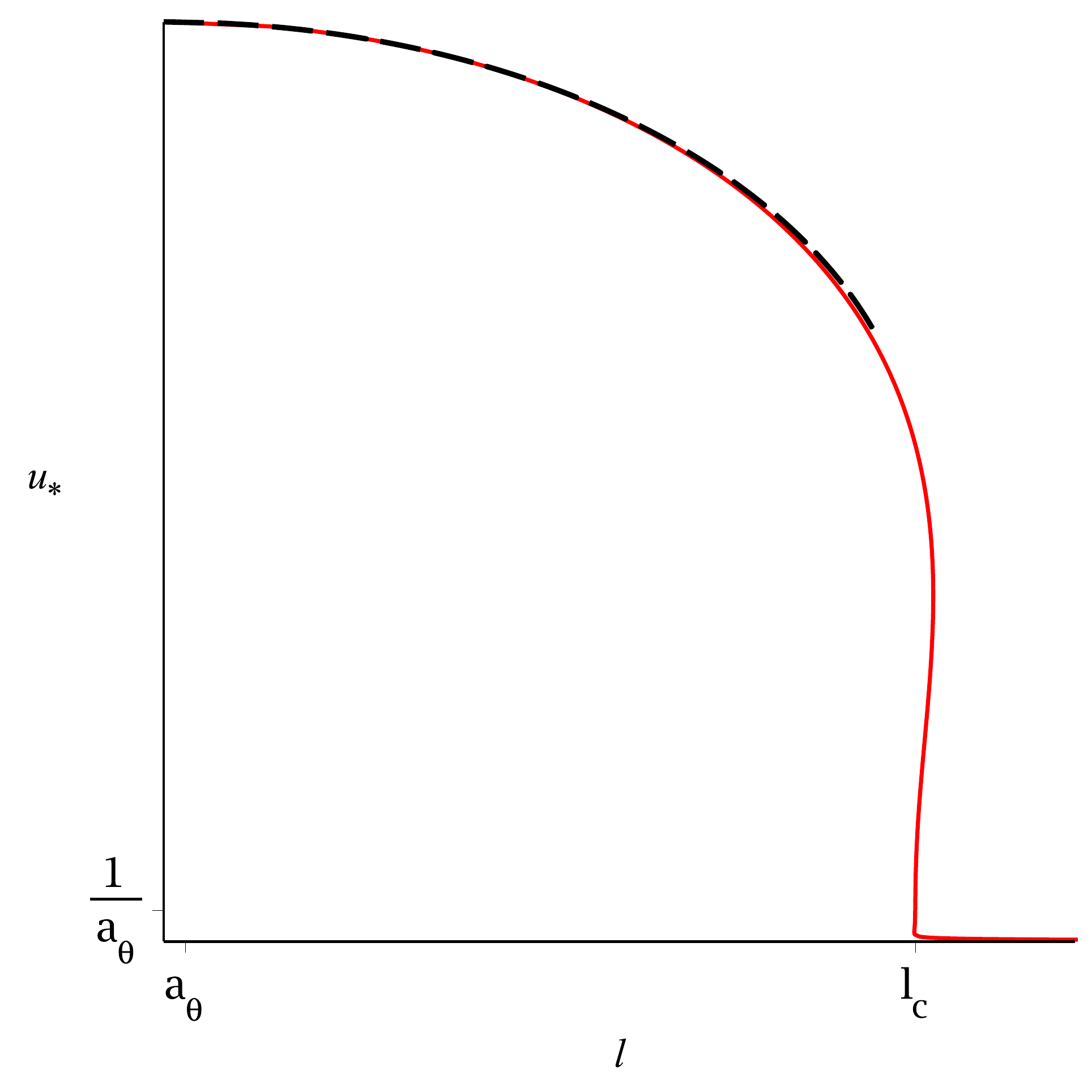}
\includegraphics[scale=0.4]{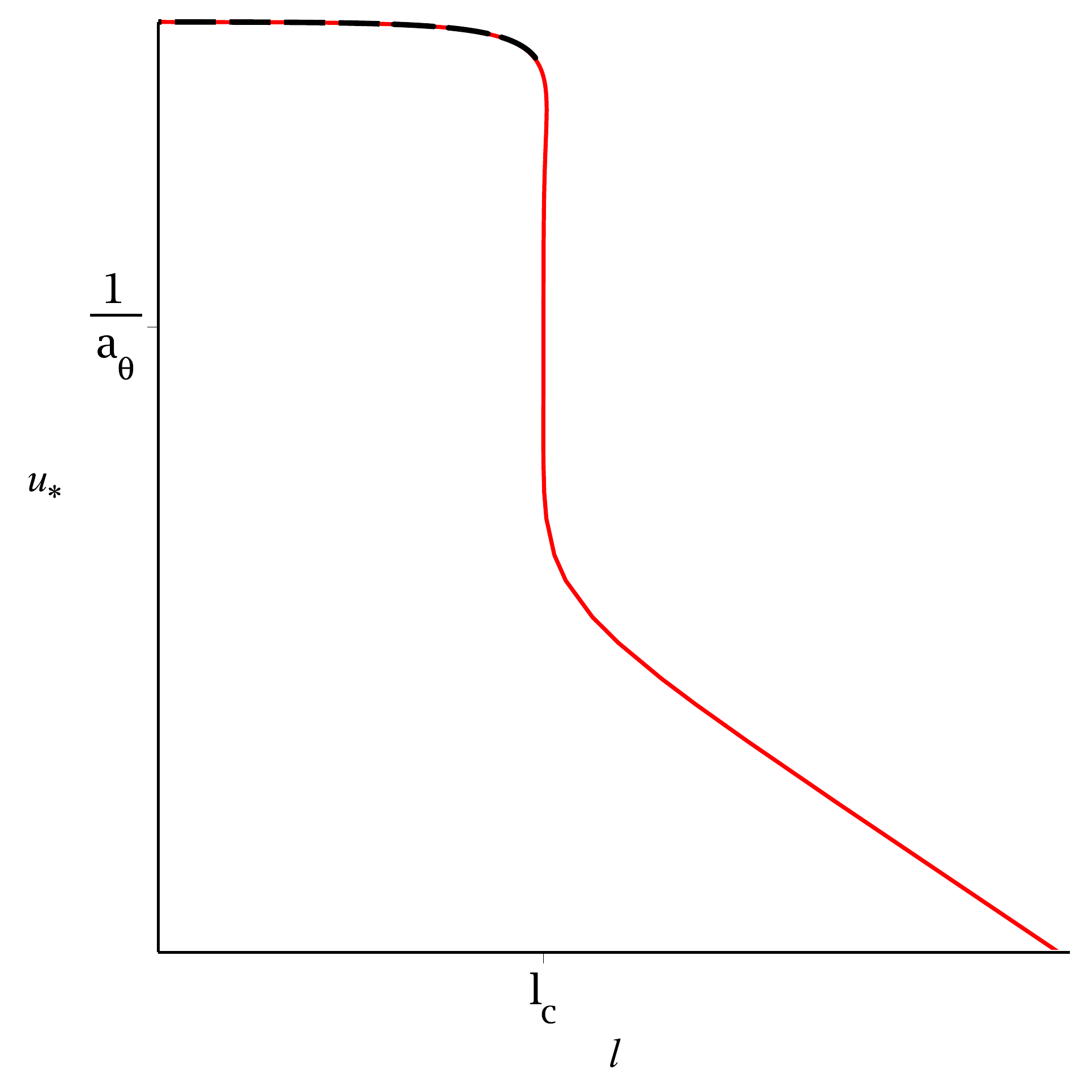}}
\caption{Point of deepest penetration $u_*$ as a function of the 
cylinder's radius $l$ for the minimal surface homologous to
a cylinder in the noncommutative theory.  
The black dashed line corresponds to equation (\ref{u-l-expansion-circle}).
Linear scale on the left, log-log scale on the right; $a_\theta u_\epsilon = 30$ for both
plots.}
\label{f8}
\end{figure}

\begin{figure}
\center{\includegraphics[scale=0.4]{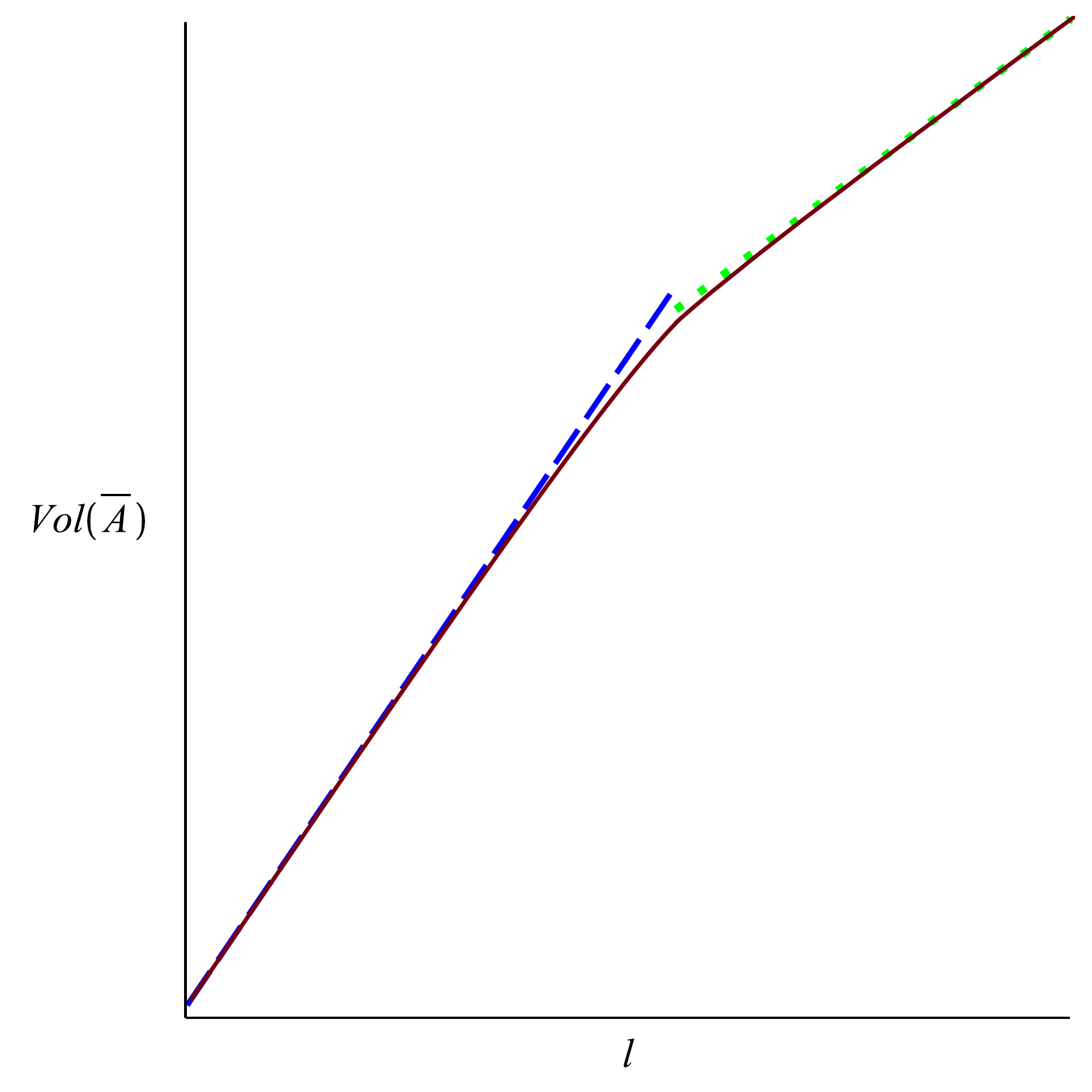}}
\caption{Area of the minimal surface homologous to
a cylinder, as a function of the cylinder's radius $l$,
with both axis shown in logarithmic scale.  
$a_\theta u_\epsilon = 30$.
The green dotted line and the blue dashed line correspond to
equations (\ref{area-l-large-circle}) and (\ref{area-l-expansion-circle}) 
respectively.}
\label{f9}
\end{figure}

Having understood the minimal surface in the large $l$ and small $l$ limits,
we now turn to $l$ near the cutoff radius $l_c =  a_\theta^2 u_\epsilon / \sqrt{3}$, 
which corresponds to  $u_* a_\theta$ close to $1$.  Figure \ref{f8} shows
the dependence of $u_*$ on $l$ over the entire range for a finite cutoff. 
We notice that near $l_c$, there are multiple values of $u_*$ at a given $l$:
just like in the case of the strip, there is a range of radii $l$ for which
there exist multiple extremal surfaces anchored on the same cylinder.  This is 
related the oscillating nature of the asymptotic solution (\ref{oscillations}).  
Since taking a large cutoff limit removes the radius $l_c$, at which
phase transition take place, to infinity, we will not attempt
a detailed study of the properties of the phase transition, which is
complicated by the oscillatory nature of the minimal surfaces near the
critical radius.  

It is interesting to notice that, apart from the details of the 
phase transition, the entanglement entropy for the cylinder
has the same qualitative behaviour as it does for the strip,
ever though the geometry of the minimal surfaces is very different.

\section{Mutual information in NCSYM}
\label{sec:mutualinfo}

\begin{figure}
\center{\includegraphics[scale=0.8]{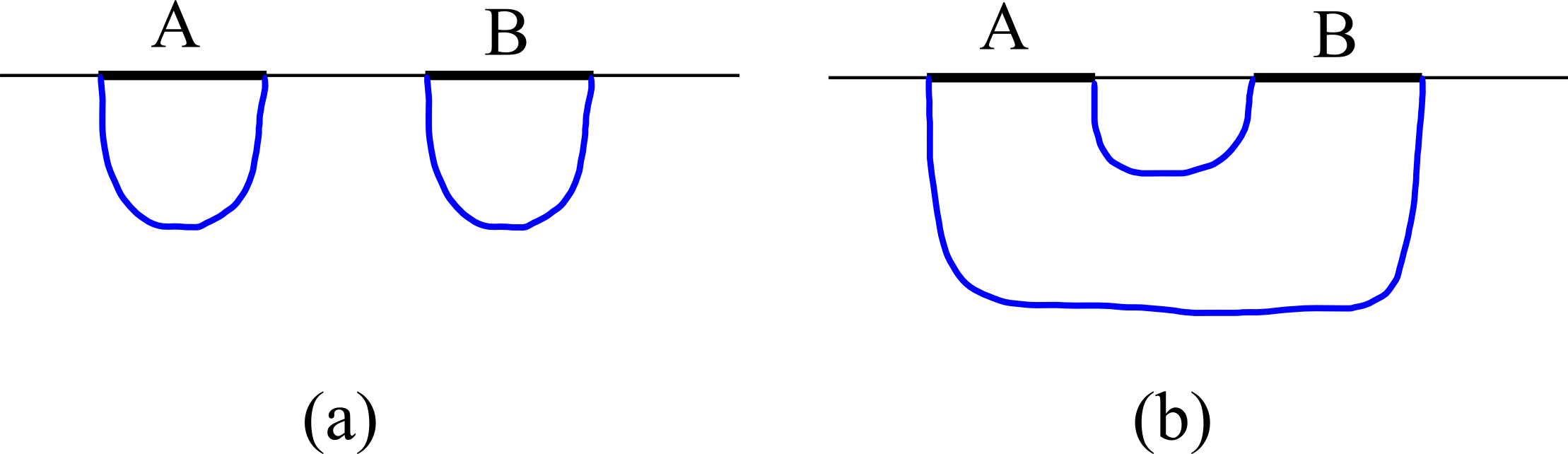}}
\caption{Minimal surface for a union of two disjoint
regions $A\cup B$ can have one of two topologies:
(a) disconnected or (b) connected.}
\label{f10}
\end{figure}

To strengthen our discussion of UV/IR mixing in noncommutative SYM theory,
it would be interesting to study the behaviour of an observable
that (in the commutative theory) is finite in the large UV cutoff limit.
One such observable is mutual information.

Consider two disjoint regions $A$ and $B$.  Mutual information is defined 
by $I(A,B) := S(A) + S(B) - S(A \cup B)$.  Subadditivity implies that mutual
information is a non-negative quantity. For local theories, holographic mutual
information is UV finite, since contributions from the 
near-boundary parts of the minimal surfaces cancels. It is known to exhibit
a phase transition \cite{Headrick:2010zt}: if the regions $A$ and $B$ have width $l$ and the distance
between them is $x$, $I(A,B)$ is nonzero for $x$ less than some $x_c$ and
zero for $x$ greater than  $x_c$, with $x_c/l$ of order 1.  The origin of this
phase transition is shown in Figure \ref{f10}: for large $x/l$, the minimal
area surface has the the topology shown in \ref{f10}(a), while for small $x/l$,
it has the topology shown in \ref{f10}(b).
Behaviour of mutual information and the existence or disappearance of this phase transition
can be used to find characteristic length scales, see for example 
\cite{Tonni:2010pv} and \cite{Fischler:2012uv}. For NCSYM we find that the mutual information 
phase transition is absent for length scales less than $l_c$.
The fact that $l_c$ depends on the UV cutoff is then a signature
of the UV/IR mixing.

To study the details of this signature, let regions $A$ and $B$
be strips of width $l_A$ and $l_B$ respectively, separated by a distance $x$.
Then, if $l_A$, $l_B$ and $x$ are held fixed as the cutoff $u_\epsilon$ is taken to infinity, 
entanglement entropies associated with strips of width $x$, $l_A$, $l_B$ and $l_A+l_B+x$ are all
extensive.  Therefore, $S(l_A)+S(l_B)  <  S(l_A+l_B+x) + S(x)$, i.e. 
the surface in Figure \ref{f10}(a)
has a smaller area than that in Figure \ref{f10}(b).  This implies that $I(A,B) = 0$
for any $x$
and there is no phase transition.  On the other hand, if $l_A$ and $l_B$ are both larger
than $l_c$, then $S(l_A) \approx S(l_B) \approx S(l_A+l_B+x)$ because to  leading order
the entanglement entropies do not depend on the width of the strip.  
Mutual information is positive (and divergent,  since entanglement entropy in the noncommutative theory does
not have a UV-finite piece) as long as
$x$ is small enough and  undergoes a phase transition as x is increased just like it does
for a local field theory.  

It would be interesting to study the behaviour of mutual information near the phase
transition in detail.  We leave this to future work.

\section{Final comments}
\label{sec:comments}

A key ingredient in our analysis was keeping the cutoff 
finite, if large.  Only when the entangling region $A$
is placed on a cutoff surface at finite $u=u_\epsilon$ can the 
correct minimal area minimal surfaces be found.  
This is especially true in the noncommutative theory,
where UV/IR mixing implies that infrared physics
is affected by the precise value of the cutoff.

We have already discussed the origins of the dependence of the 
entanglement entropy on the size (volume or area) of the 
region $A$, on the cutoff length $\epsilon$ and on the
intrinsic length scales $a_L$ and $a_\theta$ built into
our nonlocal theories.  The numerical coefficients we obtain are of
physical significance:  
In the volume law regime, the coefficient
measures whether degrees of freedom inside of $A$ are
are entangled with the outside of $A$
or with each other. Therefore,
this coefficient controls the maximum size of the 
region over which the theory thermalizes \cite{Lashkari:2013iga}.
A similar statement can be made about the 
coefficient in the area law regime.

While the open string metric gives distances in the nonlocal boundary
field theory, it is the closed string metric that
determines the causal structure of the theory. 
In a local field theory, knowledge of the density matrix $\rho_A$
in the region $A$ is enough to compute all observables
within the domain of dependence of $A$.  While we don't 
know exactly which portion of the total holographic dual
spacetime is dual to $\rho_A$ itself 
\cite{Bousso:2012sj,Czech:2012bh,Hubeny:2012wa,Bousso:2012mh},
the answer must involve the bulk  (closed string) metric and its
causal structure.  Applying this argument to our nonlocal theories,
we see that it is the bulk metric that determines
the extent of the holographic dual to the density matrix $\rho_A$.
For example, this holographic dual might 
be bounded by the minimal surface.  Then, the intersection
between the AdS boundary and the lightsheets 
originating from the minimal surface might be interpreted as 
the boundary of the ``domain of dependence"
of the region $A$ in a nonlocal theory.   We would expect that knowledge of 
the density matrix $\rho_A$ would be sufficient to determine
all observables within this ``domain of dependence''. 
This new ``domain of dependence'' is determined causally not by
the open string metric but by the bulk closed string metric
at a fixed cutoff.  This closed string metric is not isotropic,
in fact, it has a very large ``speed of light'' in the 
nonlocal directions, compared with the open string metric.  
Field theory computations show that nonlocal field theories
have large propagation speeds , see for example the
behaviour discussed in \cite{Minwalla:1999px}, or  
the observations that the propagation speed in the noncommutative theory
is effectively infinite \cite{Hashimoto:2000ys,Durhuus:2004bk}.
As a result, in a nonlocal theory the ``domain of dependence'' 
should have a very small time-like extent.
This is consistent with it being bound by lightsheets which
originate on a minimal surface which does
penetrate the bulk very far, a feature we have
observed.

A related feature of our minimal
surfaces is that they are not necessarily orthogonal
to the boundary at a finite cutoff. Therefore, for example,
the two proposals given in \cite{Hubeny:2007xt} for a 
covariant version of holographic entanglement entropy
are not necessarily equivalent, raising an interesting
question about time-dependent nonlocal theories.
Similarly,  arguments for strong subadditivity of
covariant holographic entanglement entropy in
time dependent spacetimes, in \cite{Wall:2012uf}, do not apply either
(however, the simple argument for static spacetimes,
in \cite{Headrick:2007km}, does apply, and therefore
the entanglement entropies computed in this paper
do satisfy strong subadditivity).

Since our computations were done using holography, they
are reliable in the strong coupling limit.  It would be 
interesting to see whether the same results apply at weak
coupling, with the appropriate nonlocal scale, $a_\theta$ or 
$a_L$, replaced by its weak coupling counterpart, $\sqrt \theta$
or $L$ respectively.  This might not necessarily be the case:
for example, the enhancement to the rate of dissipation provided
by noncommutativity at strong coupling is not seen at weak
coupling \cite{Edalati:2012jj}.  The analysis in \cite{Edalati:2012jj} points towards
strong coupling being necessary for scrambling in
noncommutative theory, and, if the results in \cite{Lashkari:2013iga}
can be extended to this situation, strong coupling
being necessary for extensive entanglement entropy. 
It would be interesting to settle this question by a direct computation 
of geometric entanglement entropy in a weakly coupled noncommutative theory.
Unfortunatelly, it will not be possible to learn anything from
free noncommutative theories as these are equivalent to
their commutative counterparts.

A simple example of a nonlocal field theory with volume scaling
of its entanglement entropy was given in \cite{Li:2010dr}.
In that work, it was proposed that volume scaling was
a necessary feature of entanglement entropy in a hypothetical
field theory dual to flat space.  In contrast to this
hypothetical theory, our nonlocal theories do not
have vanishing correlation functions.

Finally, it would be interesting to study other extremal surfaces 
in holographic duals to nonlocal theories, following 
the work for local theories \cite{Hubeny:2012ry},
as well as to extend our results to finite temperature.

%%%%%%%%%%%%%%%%%%%%%%%%%%%%%%%%%%%%%%%%%%%%%%%%%%%%%%%%%%%%%%%%%%%%
%  END MATTER: BIBLIOGRAPHY, ACKNOWLEDGMENTS, ...                  %
%%%%%%%%%%%%%%%%%%%%%%%%%%%%%%%%%%%%%%%%%%%%%%%%%%%%%%%%%%%%%%%%%%%%

\section*{Acknowledgments}

We are grateful for helpful discussions with Keshav Dasgupta,
Ori Ganor, Nima Lashkari, Shunji Matsuura and Mark van Raamsdonk.
This work was completed with support from the Natural Sciences and Engineering Council
of Canada (NSERC).

\bibliographystyle{JHEP}
\bibliography{new}

\end{document}